\documentclass[fleqn,usenatbib]{mnras}

\usepackage{newtxtext,newtxmath}

\usepackage[T1]{fontenc}
\usepackage{ae,aecompl}

\usepackage{graphicx}	
\usepackage{amsmath}	
\usepackage{amssymb}	
\usepackage{graphicx}	
\usepackage{amsmath}	
\usepackage{amssymb}	
\usepackage{multicol}        
\usepackage{bm}		
\usepackage{pdflscape}	
\usepackage[]{units}

\defcitealias{grudic:2016.sfe}{Paper~I}
\defcitealias{Elson:1987.ymc.profile}{EFF}

\newcommand{\be}{\begin{equation}}
\newcommand{\ee}{\end{equation}}

\newcommand{\msun}{M_{\sun}}

\usepackage[normalem]{ulem}

\title[Star Cluster Structure from Hierarchical Star Formation]{From the Top Down and Back Up Again: Star Cluster Structure from Hierarchical Star Formation}

\author[M. Grudi\'{c} et al.]{Michael Y. Grudi\'{c},$^{1}$\thanks{Contact e-mail: \href{mailto:mgrudich@caltech.edu}{mgrudich@caltech.edu}}
D\'avid Guszejnov,$^{1}$
Philip F. Hopkins,$^{1}$
Astrid Lamberts,$^{1}$\newauthor
Michael Boylan-Kolchin,$^{2}$
Norman Murray,$^{3,4}$
and Denise Schmitz$^{1}$
\\
$^{1}$TAPIR, Mailcode 350-17, California Institute of Technology, Pasadena, CA 91125, USA \\
$^{2}$Department of Astronomy, The University of Texas at Austin, 2515 Speedway, Stop C1400, Austin, TX 78712, USA \\
$^{3}$Canadian Institute for Theoretical Astrophysics, 60 St. George Street, University of Toronto, ON M5S 3H8, Canada\\
$^{4}$Canada Research Chair in Astrophysics \\
}

\pubyear{2017}

\begin{document}
\label{firstpage}
\pagerange{\pageref{firstpage}--\pageref{lastpage}}
\maketitle

\begin{abstract}
Young massive star clusters spanning $\sim 10^4 - 10^8 \msun$ in mass have been observed to have similar surface brightness profiles. We show that recent hydrodynamical simulations of star cluster formation have also produced star clusters with this structure. We argue analytically that this type of mass distribution arises naturally in the relaxation from a hierarchically-clustered distribution of stars into a monolithic star cluster through hierarchical merging. We show that initial profiles of finite maximum density will tend to produce successively shallower power-law profiles under hierarchical merging, owing to certain conservation constraints on the phase-space distribution. We perform $N$-body simulations of a pairwise merger of model star clusters and find that mergers readily produce the shallow surface brightness profiles observed in young massive clusters. Finally, we simulate the relaxation of a hierarchically-clustered mass distribution constructed from an idealized fragmentation model. Assuming only power-law spatial and kinematic scaling relations, these numerical experiments are able to reproduce the surface density profiles of observed young massive star clusters. Thus we bolster the physical motivation for the structure of young massive clusters within the paradigm of hierarchical star formation. This could have important implications for the structure and dynamics of nascent globular clusters.
\end{abstract}




\section{Introduction}
Most stars in the Universe are field stars, gravitationally bound only to their host galaxies and not to any discernible smaller element of structure. However, when the locations of initial star formation are considered, there is strong evidence that most stars are born in a statistically clustered, correlated configuration \citep{lada:2003.embedded.cluster.review,mckee:2007.review, bressert:2010.clustered.sf,gouliermis:2015.hierarchical.clustering,grasha:2017.hierarchical.clustering, gouliermis:2018.review}. The star formation efficiency $\frac{M_\star}{M_{gas}}$ of typical giant molecular clouds is only of order $1-10\%$ \citep{myers:1986.gmcs,mooney.solomon:1988,williams.mckee:1997, evans:2009.sfe,lada:2010.gmcs,heiderman:2010.gmcs,murray:2010.sfe.mw.gmc,kennicutt:2012.review,lee:2016.gmc.eff}, possibly due to stellar feedback disrupting the molecular cloud once a certain stellar mass has formed \citep{murray:molcloud.disrupt.by.rad.pressure,hopkins:fb.ism.prop,hopkins:2013.fire,grudic:2016.sfe}. The loss of binding energy from the blowout of the remaining gas can unbind the initial stellar distribution \citep{tutukov:1978,hills:1980,mathieu:1983,lada:1984,1985ApJ...294..523E,baumgardt.kroupa:2007,parmentier:2008}, allowing most or all stars to disperse into the surrounding galaxy. However, the existence of young, apparently well-relaxed star clusters within the Milky Way \citep{portegies-zwart:2010.starcluster.review} suggests that a certain fraction of star formation does lead to bound cluster formation, even in Milky Way-like conditions. In many cases, young star clusters have not had time to evolve under the effects of evaporation, dynamical relaxation, and stellar evolution, so their structures should contain some information about their initial formation. A successful model of star cluster formation will be able to clarify this relationship.

\begin{figure*}
\includegraphics[width=\textwidth]{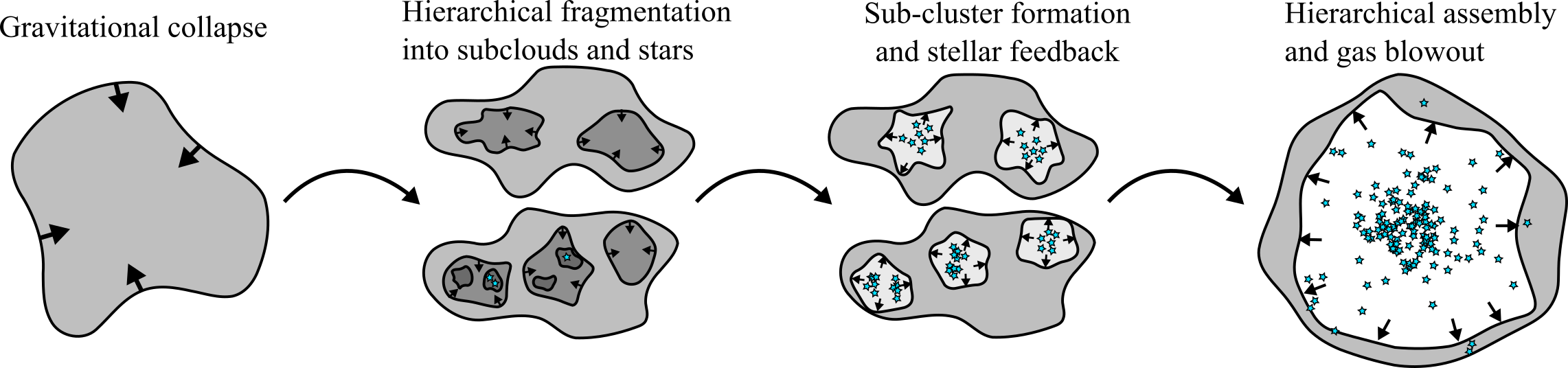}
\caption{Proposed model of cluster formation from hierarchical star formation. {\it Far left:} An unstable molecular cloud undergoes gravitational collapse. {\it Centre left}: The gravitational instability causes hierarchical fragmentation, producing a hierarchy of sub-clouds that eventually fragment into individual stars. {\it Centre right:} Stars that fragmented out of the same sub-clouds form in sub-clusters. Feedback from massive stars starts to evacuate gas locally. {\it Far right:} The sub-clusters merge hierarchically into a single cluster as stellar feedback blows out any remaining gas.}
\label{fig:cartoon}
\end{figure*}

In this paper, we discuss the formation of young massive star clusters (YMCs): star clusters that are younger than $\sim \unit[100]{Myr}$ and more massive than $\unit[10^4]{\msun}$ \citep{portegies-zwart:2010.starcluster.review}\footnote{The definition of \citet{portegies-zwart:2010.starcluster.review} also implcitly includes gravitational boundedness, however we emphasize that the observed YMCs we refer to in this text are not necessarily bound.}. Unlike mature globular clusters, which are generally well-fit by tidally-truncated models such the \citet{king:profile} profile, YMCs have been found to have extended power-law profiles with no discernible truncation, and hence are better fit by the \citet{Elson:1987.ymc.profile} surface brightness model (hereafter \citetalias{Elson:1987.ymc.profile}). This model consists of a core of finite surface brightness $\mu_0$ with an outer surface brightness profile that decays as $\mu \propto R^{-\gamma}$, where $\gamma$ is the parameter determining the logarithmic slope of the surface brightness profile, hereafter referred to as the ``profile slope''. If $\gamma \leq 2$, the integrated stellar mass is divergent, so  \citetalias{Elson:1987.ymc.profile} profiles with $\gamma \sim 2$ are referred to as ``shallow'', and have a greater proportion of their light in the power-law portion of the surface brightness profile compared to steeper profiles.

YMCs quite often do have shallow profile slopes with $\gamma$ typically ranging from 2.2 to 3.2  \citep{Elson:1987.ymc.profile,mackey:2003.ymc.profiles.lmc,mackey:2003.ymc.profiles.smc,portegies-zwart:2010.starcluster.review,ryon:2015.m83.clusters}, which correspond to 3D density profiles $\rho \propto r^{-3.2}-r^{-4.2}$ in the outer regions. The super star clusters (SSCs) of NGC 7252, despite being three to four orders of magnitude more massive than YMCs of the Local Group, also have profile slopes in this range \citep{bastian:2013.ngc7252.clusters}. This agreement across mass scales suggests some scale-free physical mechanism of bound star cluster formation, such that a shallow  \citetalias{Elson:1987.ymc.profile}-like surface brightness profile is generally produced.

One might suppose that the shallow power-law profile of young clusters somehow reflects the initial stellar configuration at the time of star formation, and a smooth cloud of gas turns into a structureless star cluster (e.g. \citealt{goodwin:1998}). However, observations and simulations \citep{maclow:2004,mckee:2007.review,kruijssen:2013.review,krumholz:2014.feedback.review} of star-forming clouds agree that the initial distribution of stellar positions in a star cluster is clumpy and hierarchical, not smooth and monolithic. Thus, presently-observed smoothly-distributed star clusters are likely to have assembled from a hierarchy of sub-clusters that fragmented out of the parent molecular cloud. If so, the present-day structure of young star clusters is the direct result of top-down fragmentation into stars followed by bottom-up assembly into a single star cluster (see Figure \ref{fig:cartoon}). In this work we investigate this physical process, arriving at an explanation for the observed structure of YMCs.

This paper is structured as follows: in Section \ref{sec:gammadist} we review observations of the structure of YMCs and compare them to the catalogue of star clusters formed in the \citet{grudic:2016.sfe} (hereafter \citetalias{grudic:2016.sfe}) suite of star cluster formation simulations. We argue that the profile slopes of YMCs are established early in a cluster's lifetime, and hence must emerge from their hierarchical formation events. In \ref{sec:analytic} we discuss how this happens, arguing analytically that the hierarchical merging of sub-clusters generally creates clusters with shallower power-law slopes through phase-space mixing. In Section \ref{sec:Nbody}, we test our analytic predictions against $N$-body simulations of collisionless pairwise star cluster mergers and the collisionless relaxation of a hierarchically-clustered mass distibution. In Section \ref{sec:discussion} we discuss various possible implications and generalizations of our results, and in Section \ref{sec:conclusion} we summarize our main results. Appendix \ref{appendix:clusterfinding} describes our algorithm for identifying bound star cluster membership from $N$-body particle data in the simulations of  \citetalias{grudic:2016.sfe}. In Appendix \ref{appendix:EFF} we derive, plot, and provide approximations for various functions that are useful in the analysis of a \citetalias{Elson:1987.ymc.profile} star cluster model in collisionless equilibrium with arbitrary $\gamma$.


\section{Profile Slopes of YMC Populations}  \label{sec:gammadist}
\begin{figure*}
\includegraphics[width=\textwidth]{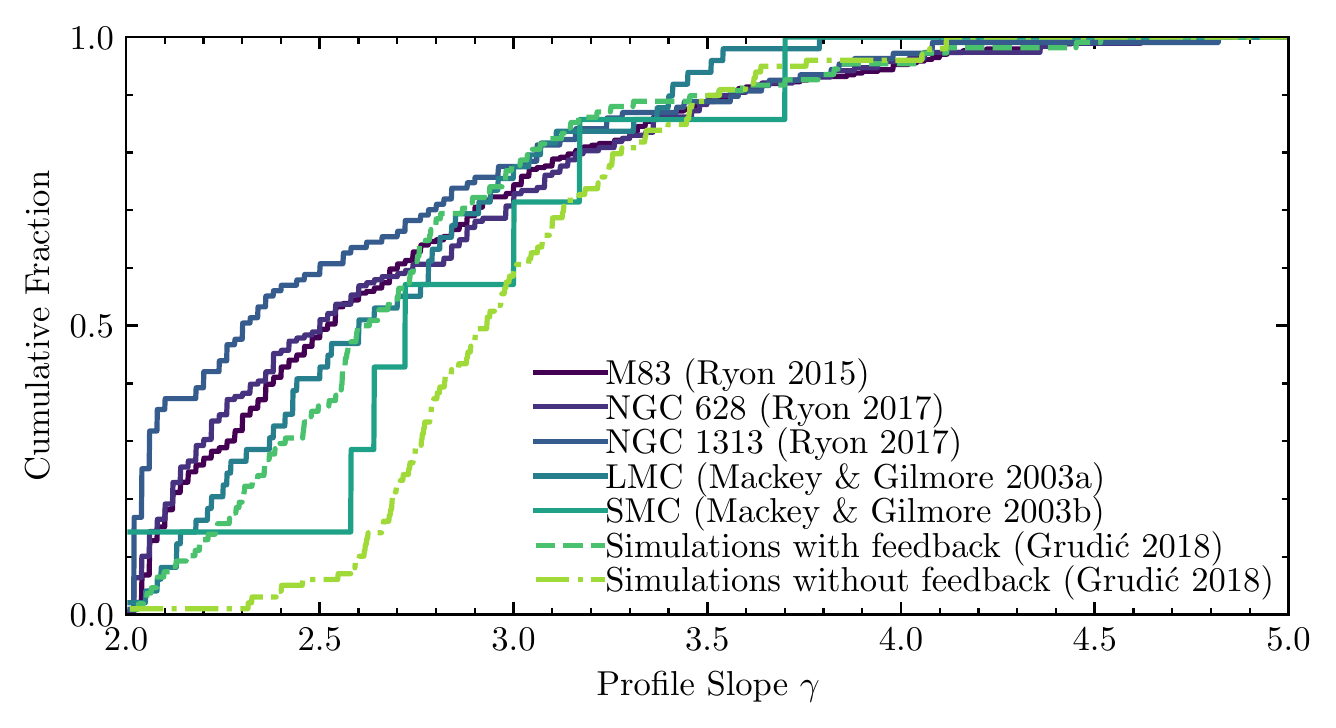}
\caption{{\it Solid:} Cumulative distribution of star cluster profile slope in the YMC populations of M83 \citep{ryon:2015.m83.clusters}, NGC 628, NGC 1313 \citep{ryon:2017.ymc.profiles}, and the Small and Large Magellanic Clouds \citep{mackey:2003.ymc.profiles.smc,mackey:2003.ymc.profiles.lmc}. {\it Dashed:} CDF for the star cluster population extracted from the simulations of \citetalias{grudic:2016.sfe}, with and without stellar feedback. For both real and simulated cluster populations, we include only those clusters that have $\gamma>2$,  as in \citet{ryon:2015.m83.clusters}. Agreement between the observed populations is quite good, however the simulations without feedback appear to have a deficit of  shallow clusters. This may be due to the greater compactness of star clusters produced in absence of feedback, which decreases the cross section for the dynamical interactions that lead to shallower profiles.}
\label{fig:gamma_cdf}
\end{figure*}

The \citetalias{Elson:1987.ymc.profile} surface brightness model commonly used to fit YMCs has the form
\begin{equation}
\mu (R) = \mu_{max} \left( 1 + \frac{R^2}{a^2}\right)^{-\gamma/2},
\label{eq:EFF}
\end{equation}
where $\mu_{max}$ is the central surface brightness, $R$ is the projected distance from the centre, $a$ is a scale radius, and $\gamma$ gives power law index of the outer brightness profile, hereafter referred to as the ``profile slope''. The corresponding 3D density profile assuming a constant mass-to-light ratio is
\begin{equation}
\rho\left(r\right) = \rho_0 \left(1 + \frac{r^2}{a^2}\right)^{-\frac{\gamma+1}{2}},
\label{eq:EFF3D}
\end{equation}
where
\begin{equation}
\rho_0 =\frac{M}{a^3} \frac{\Gamma \left(\frac{\gamma +1}{2}\right)}{\pi ^{3/2} \Gamma
   \left(\frac{\gamma -2}{2}\right)}
\end{equation}
is the central density, $M$ the total mass, $a$ the scale radius, and $\gamma$ the profile slope. This density profile can be recognized as a generalization of the \citet{plummer} model (corresponding to $\gamma=4$) to arbitrary profile slope.

Several observed YMC populations are rich enough to be able to discern an underlying distribution of profile slopes. In Figure \ref{fig:gamma_cdf} we plot the distribution of $\gamma$ as measured by  \citet{ryon:2015.m83.clusters} for YMCs in M83, \citep{ryon:2017.ymc.profiles} for NGC 628 and NGC 131, and \citet{mackey:2003.ymc.profiles.lmc,mackey:2003.ymc.profiles.smc} for the Magellanic Clouds. These clusters range from $\sim \unit[10^6-10^8]{yr}$ in age and $\sim \unit[10^4-10^6]{\msun}$ in mass. In all three populations, the median $\gamma$ is around $2.5$. In general, agreement between the observed distributions is quite good, suggesting that a population of  \citetalias{Elson:1987.ymc.profile}-like clusters with this $\gamma$ distribution arises from some common underlying physical process.

Power-law density profiles have been proposed to emerge in star clusters in various ways. A power law density profile is the hallmark of gravothermal core collapse, but an inner density profile of $\rho \propto r^{-2.2}$ should generally result \citep{lynden-bell:1980.core.collapse, cohn:1980.core.collapse}, which is unlike the outer power-law profile $\rho \propto r^{-3.5}$ typically observed in YMCs. \citet{hoerner:1957} and \citet{henon:1964} found that a $\rho\left(r\right) \propto r^{-4}$ (hence $\gamma=3$) density profile results when a uniform collisionless sphere with a Maxwellian velocity distribution undergoes violent relaxation toward collisionless equilibrium. More generally, it results from a discontinuity in the distribution of stellar mass in energy space across the boundary between bound and free orbits, as is caused by the escape of stars with positive energy after a violent relaxation event \citep{aguilar:1986,jaffe:1987, merritt:1989.violent.relaxation}. As such, this may be a good model of the initial relaxation of the smallest bound sub-structures, or at the resolution limit in star cluster formation simulations that do not resolve individual stars \citepalias[e.g.][]{grudic:2016.sfe}. However, it does not explain the fact that the majority of star clusters have $\gamma<3$.

\citet{Elson:1987.ymc.profile} suggested that the typically observed value $\gamma \sim 2.5$ corresponds to the $\rho \propto r^{-3.5}$ profile found in \citet{spitzer:1972.halo} as a steady-state solution for the outer halo of a star cluster with an inner core, but they proceeded to point out that this structure would have to be established on the two-body relaxation timescale \citep{spitzer:1987,portegies-zwart:2010.starcluster.review}:
\begin{equation}
t_{rh} = \unit[4\times 10^7]{yr} \left(\frac{M}{10^4\msun}\right)^{1/2} \left(\frac{R_{eff}}{\unit[1]{pc}}\right)^{3/2},
\label{eq:relaxation}
\end{equation}
where $R_{eff}$ is the half-mass radius (we have also assumed here that the mean mass of a star is $\unit[0.5]{\msun}$). Many YMCs are much younger than their respective two-body relaxation timescale, so this picture is not satisfactory. 

In general, scenarios requiring more than a few $\unit[]{Myr}$ can be ruled out, as good \citetalias{Elson:1987.ymc.profile} fits appear to have been achieved for quite young star clusters. Indeed, \citet{ryon:2015.m83.clusters} found no correlation of $\gamma$ with cluster age in M83, suggesting that any secular evolutionary processes occurring within these YMCs typically takes longer than $\unit[\sim 100]{Myr}$ to have an appreciable systematic effect on the outer structure. Such young cluster have not existed long enough to experience any significant number of dynamical relaxation times or orbits around the host galaxy during which they may be tidally stripped. Thus, we will explore explanations in which $\gamma$ is established over a relatively short cluster formation timescale and then evolves only slowly. The most promising of these is the other physical explanation proposed by \citetalias{Elson:1987.ymc.profile}: dissipationless relaxation following a rapid star formation event. It was noted that simulations of the  collisionless relaxation of galaxies from a clumpy, non-equlibrium state \citep{1982MNRAS.201..939V,1984ApJ...281...13M} could reproduce the range of profile slopes observed in star clusters. We will revisit this scenario in the context of modern star formation theory.

\subsection{Simulated cluster populations}
 
To guide our analytic exploration, we consider simulations of star cluster formation. The multi-physics N-body MHD simulations of \citetalias{grudic:2016.sfe} followed the collapse of a parameter survey of unstable gas clouds with a wide range of initial conditions, e.g.,  $10-1000\,\mathrm{pc}$ in diameter and $\unit[10^2-10^4]{\msun\,pc^{-2}}$ in mean surface density. We found that the clouds form stars until a certain critical stellar surface density has been reached, sufficient to disrupt the cloud via stellar feedback, which included the combination of photoionization heating, radiation pressure, shocked stellar winds and supernova explosions, approximated numerically according to the methods developed for the FIRE project in \citet{hopkins:2013.fire,fire2}. In general, we have found that the simulations with greater star formation efficiency place end with a significant fraction of the total stellar mass in gravitationally-bound, virialized star clusters.

These star clusters form via hierarchical assembly\footnote{A visualization of the star cluster formation process can be found at \url{http://www.tapir.caltech.edu/~mgrudich/gmc.mp4}}, as has been found in previous simulations following the collapse and turbulent fragmentation of molecular clouds \citep{2000ApJS..128..287K, bonnell:2003.hierarchical}. Many small subclusters first fragment out of the molecular cloud, which them go on to merge with their neighbours, eventually building up a massive star cluster. Unlike N-body simulations of star cluster assembly that have relied upon certain assumptions about the mass-loss history of the system \citep[e.g.][]{2002MNRAS.334..156S,2005ApJ...630..879F}, the process of star cluster assembly in concert with feedback-induced mass loss is followed self-consistently by including stellar feedback physics \footnote{Unlike these works, our simulations do not resolve the motions of individual stars, however.}.

We identify and catalogue those star clusters that are both well-resolved (greater than $10^3$ particles) and gravitationally bound via the algorithm described in Appendix \ref{appendix:clusterfinding}. We have found that the surface density profiles of star clusters formed in the simulations are generally well-fit by the \citetalias{Elson:1987.ymc.profile} profile, covering a range of slope parameters. In Figure \ref{fig:gamma_cdf} we plot the distribution of slopes extracted from the star cluster populations formed in the simulations of \citetalias{grudic:2016.sfe}, both with and without stellar feedback. We find that the agreement with the observed populations is within the observational scatter for the simulations that include stellar feedback, suggesting that at least the most important physics necessary for realistic star cluster structure are accounted for in the simulations. We find no strong correlation between $\gamma$ and cluster mass, age\footnote{Note however that these simulations follow the isolated formation of star clusters, and do not follow a cluster's subsequent evolution in a galactic environment.}, or radius, in agreement with \citet{ryon:2015.m83.clusters}.

The simulations without stellar feedback  also have a significant population of shallow clusters, but there is a deficit of very shallow clusters having $\gamma < 2.5$. Without stellar feedback, the population of bound star clusters tends to be richer: more stars form overall due to the absence of a force that moderates star formation. Also, the clusters are generally denser on average due to the lack of energy input from feedback; they do not undergo dynamical expansion due to mass loss. These dense, compact clusters are much less likely per orbit to merge with their neighbours, whereas mergers are more common in simulations with feedback because the clusters undergo some amount of dynamical expansion, increasing the cross section for merging. This suggests that the formation of shallow clusters has something to do with the dynamics of the cluster assembly process.

The above simulations and observations lead us to several hypotheses about the origin of YMC mass profiles:
\begin{enumerate}
\item The distribution of profile slopes does not differ greatly between different observed or simulated cluster-forming environments, if one accounts for stellar feedback in the simulations.
\item Interactions with the galactic environment are not necessary to reproduce the observed $\gamma$ distribution, as the simulations do not include these physics.
\item Few-body interactions must play a secondary role in determining the bulk structure of the cluster, as even if the simulations were capable of resolving these effects (which they are not) they do not run for any significant fraction of a half-mass relaxation time. Structural details on the scale of individual stars, such as the stellar mass function, can be neglected in favour of a mean-field, IMF-averaged approximation over timescales much less than the two-body relaxation timescale.
\end{enumerate}

It is therefore plausible that star clusters generally form with  \citetalias{Elson:1987.ymc.profile}-like surface brightness profiles, directly from their initial relaxation from their hierarchically-clustered state.


\section{Shallow Clusters Through Merging Substructure} \label{sec:analytic}

We will now develop physical intuition for how hierarchical star formation leads to the formation of star clusters with shallow power-law profiles. Consider first the initial conditions of the problem: a gas cloud collapses and undergoes star formation. Observations of the M83 YMC population suggest that the majority of the YMCs evacuate their natal gas as soon as $\unit[2-3]{Myr}$ \citep{hollyhead:2015.m83.ymcs}, at most a few orbital times. This is also the case in the \citetalias{grudic:2016.sfe} simulations. This process of rapid star formation still has some finite duration, but we may consider an idealized model wherein the stars are formed in place instantaneously, and the system then relaxes as a dissipationless $N$-body system.

This initial arrangement of stars resulting from the fragmentation of the cloud will be hierarchically clustered (e.g. \citealt{bonnell:2003.hierarchical,gouliermis:2015.hierarchical.clustering,guszejnov:2016.correlation.function,grasha:2017.hierarchical.clustering}). This is because fragmentation will leave behind substructures of all scales from the size of the parent cloud to the scale of protostellar disks \citep{hopkins:2013.correlation.function}. The proportion of the original gas cloud that is actually converted into stars will be limited by the dynamical ejection of gas and the eventual blowout due to stellar feedback (e.g. \citealt{murray:molcloud.disrupt.by.rad.pressure,grudic:2016.sfe}), but let us assume that the cloud has high ($>50\%$) star formation efficiency, which generally leads to the formation of a bound star cluster \citep{hills:1980,elmegreen:1997.open.closed.cluster.same.mf.form}. Subclusters that fragmented from the same self-gravitating parent will then be gravitationally bound to each other on average, so once they have turned into stars they will eventually merge together under dynamical friction. The result will be a sequence of hierarchical merging: subclusters will merge with their immediate neighbours that fragmented from the same parent, then the more massive cluster will merge with its neighbour, etc (see Figure \ref{fig:cartoon}). The smallest and densest structures will merge first because their respective dynamical times are the shortest, as their orbital time will be essentially the freefall time at the mean density of their parent structure, $t_{ff}\propto \rho^\frac{-1}{2}$. 

This process is certainly complex, but the success of the \citetalias{grudic:2016.sfe} simulations in producing star clusters with the correct structure out of softened, equal-mass star particles encourages us to consider a collisionless kinetic treatment of the problem. We approximate the dynamics as those of an ensemble of stars with phase-space distribution function $f\left(\textit{\textbf{x}},\textit{\textbf{v},t}\right)$, which evolves according to the collisionless Boltzmann equation:
\begin{equation}
\frac{\mathrm{D}f}{\mathrm{D}t} = 0,
\label{eq:boltzmann}
\end{equation}
where $\frac{\mathrm{D}}{\mathrm{D}t}$ denotes the Lagrangian time derivative along the flow of the system determined by the Hamiltonian with the usual kinetic and gravitational terms. In other words, the phase-space density $f$ is conserved along trajectories of the system. Formally, this does link the initial state of a hierarchical stellar distribution to the final state of a monolithic star cluster. However, it cannot be applied directly: while the fine-grained distribution function $f$ is indeed conserved in a dissipationless relaxation process, the measurable quantity in any observation or $N$-body simulation is the coarse-grained distribution $\bar{f}$:
\begin{equation}
\bar{f}\left(\textit{\textbf{x}},\textit{\textbf{v}}, t\right) = f\left(\textit{\textbf{x}},\textit{\textbf{v}}, t\right) \ast K\left(\frac{\textit{\textbf{x}}}{\sigma_x},\frac{\textit{\textbf{v}}}{\sigma_v}\right),
\end{equation}
where $K$ is some 6-dimensional smoothing kernel, $\sigma_x$ and $\sigma_v$ are the practical resolution limits of position and velocity measurements, and $\ast$ represents phase-space convolution. In observations and $N$-body simulations, the finite masses of the bodies impose a mass scale that ultimately determines the practical limit of phase-space resolution: the support of the smoothing kernel must contain a certain number of bodies to be able to convert between the full discrete description and the continuum approximation in any meaningful way.

The collisionless Boltzmann equation does not require that $\bar{f}$ be conserved along phase-space trajectories. To the contrary, in a system relaxing violently toward equilibrium, phase-space elements of varying  $f$ tend to be stretched out and tangled together until eventually it is impossible to recover the original value of $f$ at any resolution at which the continuum limit actually applies \citep{lynden-bell67,dehnen:2005.mixing}. The result is a ``dilution'' of mass in phase-space, wherein $\bar{f}$ will generally decrease. This process is clearly essential in the relaxation of a hierarchically-clustered mass distribution into a monolithic cluster, as the the initial clumpy state contains more information than the smooth final state, so this information must be effectively lost as mixing entropy. We expect that in collisionless hierarchical cluster assembly dominated by typically equal-mass mergers, violent relaxation should be efficient at driving this phase-space dilution. 

The phase-space mixing theorem derived in \citet{dehnen:2005.mixing} makes it possible to constrain the evolution of the phase-space distribution in hierarchical merging. \citeauthor{dehnen:2005.mixing} found that when two collisionless self-gravitating systems merge, the following function of the coarse-grained phase-space density must strictly decrease for all $f$:
\begin{equation}
D\left(f\right) = \int_{\bar{f}\left(\bf{x},\bf{v}\right) > f} \left(\bar{f}\left(\bf{x},\bf{v}\right) - f\right)\,\rm{d}^3x\,\rm{d}^3v,
\end{equation}
which is known as the {\it excess mass function}. This mixing theorem was used to explain why the inner density profile of a collisionless merger product must have the same slope as the steeper of the progenitors \citep[e.g.][]{boylan-kolchin:2005,kazantzidis:2006}. It thus immediately follows that two EFF-like systems must merge into a system with a flat inner density profile \footnote{In fact, this follows intuitively from the requirement that the maximum phase-space density cannot increase. Systems in virial equilibrium with flat inner profiles have a maximum phase-space density, while systems with power-law inner profiles do not.}. 

We can also use the mixing theorem to constrain the outer density profile of the merger. For this purpose, it is more convenient to consider the {\it reciprocal} excess mass function $M-D\left(f\right)$, where $M$ is the total mass of the system; this quantity must strictly increase during mixing. \citeauthor{dehnen:2005.mixing} showed that for a system with an 3D outer density profile $\rho \propto r^{-\gamma-1}$,
\begin{equation}
M-D\left(f\right) \propto f^{\frac{2\gamma-4}{2\gamma -1}}.
\end{equation}
For values of $\gamma$ giving finite mass ($\gamma >2$), the exponent $f^{\frac{2\gamma-4}{2\gamma -1}}$ increases monotonically from 0 at $\gamma=2$ to 1 as $\gamma\rightarrow \infty$. Hence $M-D\left(f\right)$ is a steeper function of $f$ for star clusters with steeper outer profiles. Therefore, when two collisionless systems merge, the requirement that the reciprocal mass function for the whole system must increase for all $f$ implies that the function must be at least as shallow as the shallower of the two systems in isolation. Consequently, {\it the outer density profile of merger product of two collisionless systems can be no steeper than the shallower of the two progenitors}. We are thus able explain why hierarchical merging does not produce steeper density profiles than existed originally, however it remains to explain why it might drive the system toward shallower slopes.

\subsection{Similarity solution}
A shallow outer density profile profile can be associated with mass being spread over many orders of magnitude in phase space density. In particular, $\mathrm{d}M/\mathrm{d}\log \bar{f} \sim \epsilon$, where $\epsilon$ is some small fraction of the total mass of the system. More generally, if we consider any parameter describing a ``scale" that approaches 0 far away from the system, be it spatial scale, density, phase-space density, or velocity dispersion, it also holds that 
\begin{equation}
\frac{\mathrm{d}M}{\log x} \sim const.
\end{equation}
for shallow clusters, where $x$ is the chosen scale parameter. In \citet{guszejnov:2017.universal.scaling}, we argue that such a broad distribution of mass across different scales is a general feature of systems formed under the action of gravity and supersonic turbulence, whose equations can be cast in a scale-free form under the physical conditions relevant to star formation. Therefore, $\gamma \sim 2$ is the expected result of hierarchical cluster formation in the limit where the hierarchy of substructures covers a large range of scales. In both the fragmentation that produces the hierarchical structure, and the merging that effaces it, the physics can prefer no particular scale, and hence leave a small fraction of the total mass behind at each scale, hence the flat distribution of mass in $\log f$. 

This argument predicts $\gamma=2$ in the limit of cluster formation from a deep hierarchical merger tree; in effect, this is the fixed point for the outer density profile in hierarchical merging. However, clusters with $\gamma>2$ remain to be explained. Furthermore, we know that some of the simulated star clusters plotted in Figure \ref{fig:gamma_cdf} do not have particularly extended merger histories; inspection of their merger histories of the least well-resolved clusters considered generally reveals no more than $2-3$ major mergers. There is clearly some mechanism that allows clusters to reach shallow slopes with only limited merger histories, which must arise from some change in $\gamma$ in the pairwise merging of star clusters.

\subsection{Shallower density profiles through pairwise merging} \label{sec:gammaprime}
\begin{figure}
\includegraphics[width=\columnwidth]{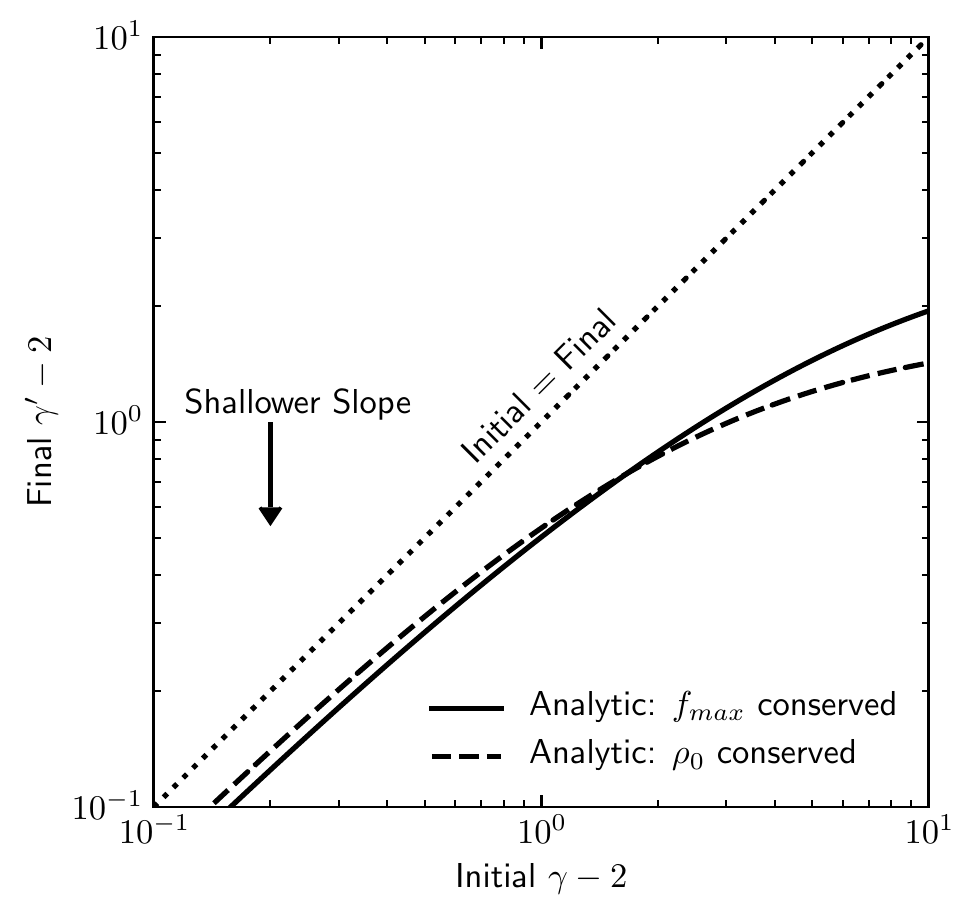}
\caption{Final surface brightness slope $\gamma'$ of the star cluster produced in a merger as a function of the initial $\gamma$ of two merging clusters with equal $\gamma$, mass, and size, assuming that the relaxed merger has an \citetalias{Elson:1987.ymc.profile} profile. We plot the analytic predictions assuming that the maximum phase-space density $f_{max}$ (solid) and the maximum density $\rho_0$ (dashed) are conserved; the two models predict similar results: merging of clusters of equal size and mass always produces a shallower profile than existed before, driving star clusters toward $\gamma=2$ regardless of their initial structure. We also plot the results of the simulated mergers described Section \ref{sec:Nbody:plummer}, which do not agree exactly with either model but predict the same overall trend of the formation of shallower profiles.}
\label{fig:gammaprime}
\end{figure}

Let us idealize hierarchical cluster formation as a sequence of pairwise cluster mergers. By symmetry, such a merger would most typically involve two clusters of similar size, mass and shape, so we will determine the outcome of a merger of identical star clusters described by EFF profiles with $M=a=1$ and a particular value of $\gamma$. Since the two clusters fragmented out of the same parent under gravitational instability, the two clusters can be expected to be gravitationally bound to each other; for simplicity we will consider the case in which they collide on a marginally-bound parabolic orbit with pericentre smaller enough for the clusters to disrupt each other in one or two passes. In a marginally-bound, collisionless merger, mass and energy are approximately conserved \citep{white:1979.mergers}, so we assume mass and energy are conserved for simplicity. Furthermore, we assume that the merger product is another star cluster with an \citetalias{Elson:1987.ymc.profile} profile with parameters $M'=2M$, $a'$ and $\gamma'$

If the merger is homologous ($\gamma'=\gamma$), mass and energy conservation imply that $M'=2$ and $a'=2$. Then the coarse-grained phase-space density $\bar{f} \propto G^{-3/2} M^{-1/2} a^{-3/2}$ in the neighbourhood of an average star is rescaled by $\frac{1}{4}$, which satisfies the constraint that $\bar{f}$ must decrease in the evolution of the system. This ``uniform mixing'' approximation has proven to be quite predictive in the case of dissipationless elliptical galaxy mergers \citep{shen:2003.galaxy.sizes, cole:2000.galform,boylan-kolchin:2005,hopkins:2009.dry.mergers}. However, the physical nature of phase-space mixing and violent relaxation in elliptical galaxy mergers may well be qualitatively different from star cluster mergers: the cusps of elliptical galaxies are scale-free, so the phase-space dilution factor tends to be roughly constant throughout the system, leading to uniform mixing. Meanwhile star clusters with flat inner profiles do have a characteristic scale imprinted by the maximum density or maximum phase-space density; some memory of the maximum density should persist in the merger.


We make the ansatz, to be justified in \S\ref{sec:Nbody:plummer}, that the maximum phase-space density persists throughout the merger, as phase mixing becomes less efficient as $f\rightarrow f_{max}$, where $f_{max}$ is the maximum phase-space density found in either cluster. If so, then $\gamma$ cannot remain the same while preserving mass and energy, as if it did then $f_{max}$ would take $\nicefrac{1}{4}$ its original value. Assuming that the merger product is an  \citetalias{Elson:1987.ymc.profile} cluster, and conservation of mass, energy and $f_{max}$, we arrive at the following equation for the final cluster's slope $\gamma'$:
\begin{equation}
\mathcal{F}\left(\gamma'\right) = 2^{5/2}\frac{\mathcal{F}\left(\gamma\right)}{ \mathcal{W}\left(\gamma\right)},
\label{eq:gammaprime}
\end{equation}
where $\mathcal{W}\left(\gamma\right)$ and $\mathcal{F}\left(\gamma\right)$ are the dimensionless functions that contain the $\gamma$ dependence of a cluster's energy and maximum phase-space density (see Equations \ref{Wapprox} and \ref{Fapprox} for approximate forms and Figures \ref{fig:energy} and \ref{fig:fmax} for plots of these functions). This equation can be solved for $\gamma'$ numerically. In the case of merging equal mass and size \citet{plummer} models ($\gamma=4$), the solution is $\gamma' = 2.83$: the final cluster is shallower than its progenitors.

We also consider the ansatz that the central density $\rho_0$ is conserved. In practice, the predictions of the two ans\"{a}tze are similar (see Figure \ref{fig:gammaprime}). In general, the models predict that $2<\gamma'<\gamma$, so a sequence of mergers will drive $\gamma$ toward a fixed point of 2. Intuitively, mass and energy conservation require the final mass and effective radius to roughly double. This must be achieved without changing the central (phase-space) density significantly, so a shallower slope is required, because a shallower cluster has greater central (phase-space) density for a given half-mass radius.

By the arguments above, even very steep ($\gamma\sim10$) clusters of similar size and mass will merge into a cluster with $\gamma\sim4$, so only $1-2$ major mergers are needed to get a cluster into the interval between 2 and 3 in which most YMCs lie (Figure \ref{fig:gamma_cdf}). As we have established that $\gamma$ must be established quite early in a cluster's lifetime, this merger history comes from the star cluster's hierarchical assembly process.

\section{N-body experiments} \label{sec:Nbody}
In the previous section, two claims were made that require verification: that the maximum phase-space density is conserved in a collisionless star cluster merger, and that the sequence of mergers necessary to produce an \citetalias{Elson:1987.ymc.profile}-like cluster with $\gamma \sim 2-3$ can arise from the relaxation of a hierarchically-clustered stellar distribution. Now we shall verify these claims with $N$-body numerical experiments, first of a sequence of pairwise mergers and then of a hierarchically-clustered configuration. We use the multi-physics code {\small GIZMO} \citep{hopkins:gizmo} in a pure $N$-body configuration. Gravity is solved with a hierarchical BH-tree algorithm derivative of {\small{GADGET-3}} \citep{springel:gadget}. We do not simulate the motion of individual stars, but rather approximate the solution of the collisionless Vlasov-Poisson equation with a Monte Carlo sampling of the distribution function with equal-mass, softened particles. Throughout, we adopt units such that $G=1$.

\subsection{Pairwise cluster mergers} \label{sec:Nbody:plummer}
\begin{table*}
\begin{tabular}{l|l||l|l|l|l|l|l|l|l|l}
\hline
Run & $M$ & $R_{eff}$ & $\gamma$ & $M'$ & $R_{eff}'$ & $\gamma'$ & Predicted $\gamma'$ & Ellipticity & $\hat{\chi}_{fit}^2$ \\
\hline
Merger 1 &      1.00 &       1.30 &       4.00 &       1.90 &       2.24 &       $2.69 \pm       0.06$ & 2.83 &  0.25 & 78.04\\ 
Merger 2 &      1.90 &       2.24 &       2.69 &       3.57 &       4.22 &       $2.48 \pm       0.03$ & 2.37 &   0.14 & 212.24\\ 
Merger 3 &      3.57 &       4.22 &       2.48 &       6.53 &       7.65 &       $2.21 \pm       0.01$ &  2.27  & 0.13 & 142.20\\
\label{table2}
\end{tabular}
\caption{Parameters and results of the sequence of simulated mergers of identical EFF-like star clusters: Initial cluster masses $M$, initial half-mass radii $R_{eff}$, initial profile slope $\gamma$, final cluster mass $M'$, final half-mass radius $R_{eff}'$, final fitted profile slope $\gamma'$,  analytically-predicted $\gamma'$ according to Equation \ref{eq:gammaprime}, cluster ellipticity, and the reduced $\chi^2$ for the fit of the final surface density profile to the EFF model. We give $\hat{\chi}_{fit}^2$ for the worst of three fits of the final cluster's surface density profile as projected in three orthogonal planes. The quoted uncertainty in $\gamma'$ includes the variation between the three different fit results.}
\label{table:pairwise.merger}
\end{table*}

\begin{figure*}
\includegraphics[width=\textwidth]{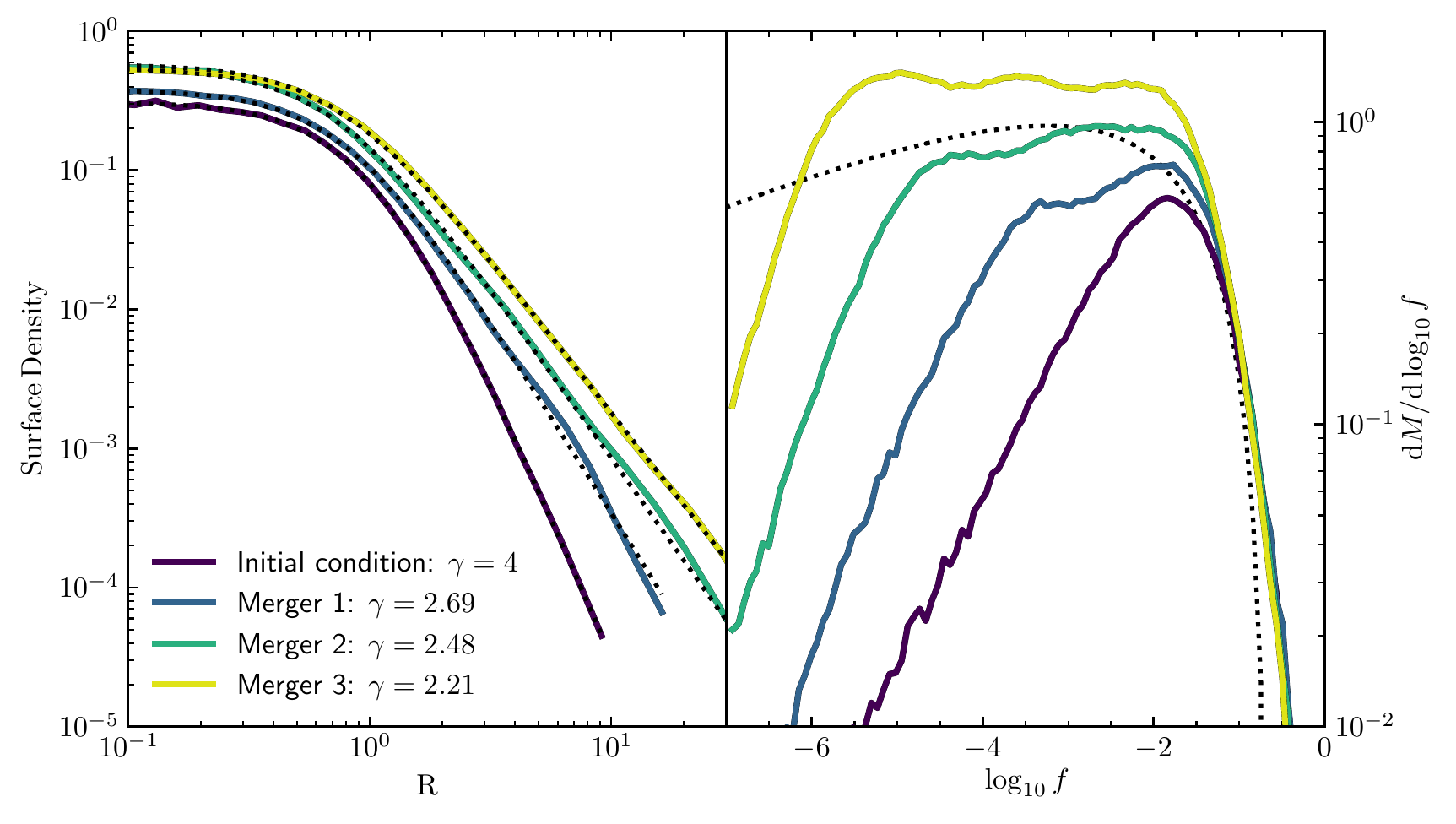}
\caption{Results of the successive pairwise merging of star clusters, starting with a pair of identical Plummer models. \textit{Left}: Cluster surface density profiles for the initial Plummer model and the three successive merger products. The mergers generally do shallow the surface density profile toward $\gamma=2$. Fits to the EFF model are shown as dotted lines. \textit{Right}: Distribution of mass in log phase-space density, $\frac{\mathrm{d}M}{\mathrm{d}\log_{10}f}$, for the simulated clusters. The mergers generally conserve the maximum phase-space density and distribute the mass across more orders of magnitude in $f$, gradually building up the flat distribution associated with shallower surface density profiles ($\gamma \sim 2$). The dotted line shows what $\frac{\mathrm{d}M}{\mathrm{d}\log_{10}f}$ would be for Merger 3 if the phase-space distribution function were that of an isotropic \citetalias{Elson:1987.ymc.profile} model with fitted parameters fitted from the surface density profile.}
\label{fig:pairwise.merger}
\end{figure*}

We first simulate the merger of two Plummer model clusters ($\gamma=4$) to test the ans\"{a}tze that their maximum phase-space density should be conserved and that the end product should be well-fit by an \citetalias{Elson:1987.ymc.profile} profile with $\gamma$ given by the solution of Equation \ref{eq:gammaprime}. Once these clusters have merged and the cluster has relaxed to a steady state, we extract this cluster, copy it, and set it up to merge with its copy. To avoid building up a spurious anisotropy along the axis of approach, the orientations of the clusters are randomized between mergers. We repeat this for a total of three simulated mergers. The Plummer-equivalent gravitational softening length is fixed at $0.1$ in all runs.
\subsubsection{Initial conditions}
We construct two Plummer cluster models in collisionless equilibrium, randomly sampling the positions of $125000$ particles per cluster according to the 3D EFF distribution (Equation \ref{eq:EFF3D}) with $M=a=1$ and $\gamma=4$. The velocity distribution is assumed to be isotropic and is randomly sampled according to the phase-space distribution function of Equation \ref{eq:feapprox}, which is exact for the Plummer model. We find that a single such cluster evolved in isolation for $10^4$ half-mass dynamical times has no significant evolution from the Plummer model, so we expect that the particle number is sufficient so that collisional effects play no major role in the merger, which happens after $\sim300$ dynamical times. We place the cluster centres 100 length units from each other, with the relative velocity adjusted for a parabolic encounter with a pericentric radius of 1.6, which is just close enough that the clusters merge in a single pass. We set up the two subsequent mergers in the same way, but we scale the pericentric radius to the half-mass radius of the cluster. 

\subsubsection{Results}
In all simulations, the clusters approach and merge in a single pass after $\mathcal{O}(10^2)$ time units, and by the end of the simulation at $t=1000$ the new cluster has approached a new collisionless equilibrium. A fraction of the particles are ejected from the system, so the assumption that the final cluster will contain all initial mass and energy does not hold exactly, but the fraction is always $<10\%$. Free particles are deleted from subsequent merger simulations.

Data on the formed clusters are presented in Table \ref{table:pairwise.merger}. We perform \citetalias{Elson:1987.ymc.profile} fits on the final surface density profiles as projected in three orthogonal different planes. The particle positions are binned into annuli around the centre of the cluster, and we fit the masses within each bin to the EFF model via $\chi^2$ minimization. Since we interpret the particle states as a Monte Carlo sampling of the phase-space distribution, the uncertainty of the mass $m$ in each bin is taken to be the Poisson sampling error $\frac{m}{\sqrt{N}}$, where $N$ is the number of particles in the bin (valid for sufficiently large $N$). We find that the EFF model always fits the surface density profiles reasonably well (Figure \ref{fig:pairwise.merger}, panel 1), but not exactly; the reduced $\chi^2$ of the fits are on the order of 100. The clusters are only weakly triaxial, with ellipticity 0.25 at most, so the fit results from different projection planes do not vary greatly. Mergers 2 and 3 both reduce the ellipticity initially created by Merger 1.

We find that the successive mergers do shallow the surface density profiles (Figure \ref{fig:pairwise.merger} clusters with $\gamma=4$ merge into $\gamma=2.69$, then $2.69$ into $2.48$, and then $2.48$ into $2.21$. This is not in exact agreement with the analytic predictions of Section \ref{sec:gammaprime} assuming either conservation of density or phase-space density, however the analytic and numerical predictions of $\gamma$ agree to within $0.1$, and agree upon the general trend of a decrease toward $\gamma=2$. Perfect agreement with the model is not expected because of the many approximations we have invoked. In particular, it is likely that the obtained slope of $2.69$ is shallower than the predicted $2.8$ due to the fact that the merger orbit had non-zero angular momentum, which must be redistributed in the final configuration. This would give a mass distribution that is more extended (ie. with a shallower slope) than a cluster of equal energy with no net angular momentum.

The last assumption of Section \ref{sec:gammaprime} to be verified is conservation of the maximum phase-space density. We estimate the coarse-grained phase-space density in the neighbourhood of particle $i$ in the most straightforward way, generally known as the pseudo-phase-space density \citep{navarro:2001.pseudo.phase.space}:
\begin{equation}
\bar{f_i} \propto \frac{\rho_i}{\sigma_i^3},
\label{eq:pseudo.phase.space}
\end{equation}
where $\rho_i=\frac{m_i}{V_i}$ is the density of the particle estimated from its effective volume \citep{hopkins:gizmo}, and $\sigma_i$ is the local velocity dispersion computed from the velocities of the particle's 32 nearest neighbours. \footnote{Much more accurate estimates of $\bar{f}$ from $N$-body data exist \citep{arad:2004.phase.space.density,ascasibar:2005.phase.space.density}, but the pseudo-phase-space density is suitable for the purposes of this limited analysis.} In Figure \ref{fig:pairwise.merger}, panel 2 we plot the distribution  $\frac{\mathrm{d}M}{\mathrm{d}\log f}$ and find that indeed, the maximum phase space density (corresponding to the upper cutoff of the distribution) is conserved from the initial Plummer model to the final merger. Thus, the deviation of $\gamma$ from analytic predictions is due to the deviation of the phase-space distribution of the cluster from from that of an isotropic EFF model. This is evident in Figure \ref{fig:pairwise.merger}: despite the good apparent fits of the surface density of Merger 3 to the EFF model, its distribution of phase space densities looks quite different from that of an isotropic EFF model in collisionless equilibrium (shown as the dotted line). Rather than having the predicted asymptotic $\propto f^{\frac{2\gamma-4}{2\gamma-1}}$ dependence for small $f$, the distribution is flat over a finite interval, then falls off steeply above and below that interval. The phase-space density at the lower cutoff corresponds to the mean phase-space density of particles near 100 distance units from the cluster centre, which is the initial separation between the clusters in the merger setup and hence where we expect any scale-free behaviour to break down. 

From these results we may conclude that the assumptions of Section \ref{sec:gammaprime} were largely valid: the collisionless merger of two  \citetalias{Elson:1987.ymc.profile} clusters fits reasonably well to another \citetalias{Elson:1987.ymc.profile} cluster, at least in its surface density profile. The profile slope $\gamma$ is close to that analytically determined by conservation of mass, energy and $f_{max}$; conservation of mass and energy hold approximately, while conservation of $f_{max}$ holds exactly, to the extent that can be tested by our noisy estimate of the phase-space density.

\subsection{Relaxation of a Hierarchically-Clustered Mass Distribution} \label{sec:Nbody:hierarchical}
Now we wish to examine whether a hierarchically-clustered distribution of stars with realistic spatial and kinematic scaling relations can form an \citetalias{Elson:1987.ymc.profile}-like star cluster as it relaxes toward collisionless equilibrium. We arrange particles in such a configuration and simulate their dynamical evolution from the hierarchically-clustered state.
\subsubsection{Initial Conditions} \label{sec:Nbody:hierarchical.ics}
We initialize $64^3$ particles in a hierarchically-fragmented configuration by recursively bifurcating a population of subclusters, starting with a single cluster of unit mass centred at the origin. In each bifurcation, the mass ratio $q$ of the two child fragments is sampled from the log-normal distribution \footnote{The choice of a lognormal mass ratio distribution was arbitrary; we have also run simulations where $q$ is always 1, and have found no major difference in our results.} with $\langle q \rangle = 1$ and $\sigma_{\log q}=1$. The masses of the fragments are then
\begin{equation}
\begin{split}
m_1 &= \frac{q}{1+q} m_{parent}, \\ 
m_2 &= \frac{1}{1+q} m_{parent}.
\label{eq:mi}
\end{split}
\end{equation}
The relative separation of the fragments $\Delta \textit{\textbf{x}}$ is sampled from a 3D normal distribution with variance $\sigma_{x}^2$. We scale $\sigma_{x}^2$ to achieve the desired  two-point spatial correlation function $\xi\left(r\right) \propto r^{-2}$, where
\begin{equation}
1 + \xi\left(r\right) = \frac{\langle n\left(r\right)\rangle}{\langle n \rangle},
\end{equation}
is the ratio between the average number density of particles in a spherical shell of radius $r$ around a star to the mean stellar number density of the system. $\xi\left(r\right)$ quantifies the tightness of the hierarchical clustering at a given scale $r$. The form $\xi\left(r\right) \propto r^{-2}$ matches observations of young star clusters on scales greater than $\unit[0.01]{pc}$, and is predicted by numerical simulations and general considerations of the scale-free interplay of gravity and supersonic turbulence \citep{guszejnov:2016.correlation.function,guszejnov:2017.universal.scaling}. This scaling is achieved by the ``isothermal'' scaling $\sigma_x \propto m_{parent}$, so $\sigma_x$ is thus determined down to a constant scale factor.

With the separation $\Delta \textit{\textbf{x}}$ thus sampled, the child clusters are displaced so as to preserve the centre of mass:
\begin{equation}
\begin{split}
\textit{\textbf{x}}_1 &= \textit{\textbf{x}}_{parent} + \frac{1}{1+q} \Delta \textit{\textbf{x}}, \\
\textit{\textbf{x}}_2 &= \textit{\textbf{x}}_{parent} - \frac{q}{1+q} \Delta \textit{\textbf{x}}.
\label{eq:xi}
\end{split}
\end{equation}

Lastly, the relative velocity $\Delta \textit{\textbf{v}}$ of the child clusters is sampled from a 3D normal distribution scaled to emulate the $v^2 \propto R$ kinematic relation of that is generally observed in GMCs \citep{larson:gmc.scalings, solomon:gmc.scalings,bolatto:2008.gmc.properties} and is robustly reproduced in simulations of isothermal, self-gravitating turbulent clouds \citep{kritsuk:2013.larson.laws}, the idea being that protostars will inherit the kinematics of the ISM from which they formed. This scaling relation is achieved by setting $\sigma_v^2 \propto M^{4/3}$. Then, to conserve momentum,
\begin{equation}
\begin{split}
\textit{\textbf{v}}_1 &= \textit{\textbf{v}}_{parent} + \frac{1}{1+q} \Delta \textit{\textbf{v}}, \\ 
\textit{\textbf{v}}_2 &= \textit{\textbf{v}}_{parent} - \frac{q}{1+q} \Delta \textit{\textbf{v}}.
\label{eq:vi}
\end{split}
\end{equation}

The bifurcation iteration described by equations \ref{eq:mi} to \ref{eq:vi} is applied recursively until the mass of a single particle is reached, so structures exist on all mass scales down to the mass of individual particles. However, recall that these $N$-body simulations of equal-mass, softened particles are to be interpreted as a Monte Carlo approximation of the solution of the collisionless Boltzmann equation. For this to be valid, any resolved structures should be sampled by a certain number of particles, as biases in the dynamics due to the discreteness of the particles are not part of the desired solution. For this reason, once the clustered configuration has been generated, we smooth the initial conditions by displacing each particle by a random normally-distributed offset with $\sigma= 10^{-3}$; this ensures that structures in the initial conditions are sampled by at least $\sim100$ particles. We also set the Plummer-equivalent gravitational softening length to $10^{-3}$ for consistency \citep[e.g.][]{barnes:2012.softening.is.smoothing}.

\begin{figure*}
\includegraphics[width=\textwidth]{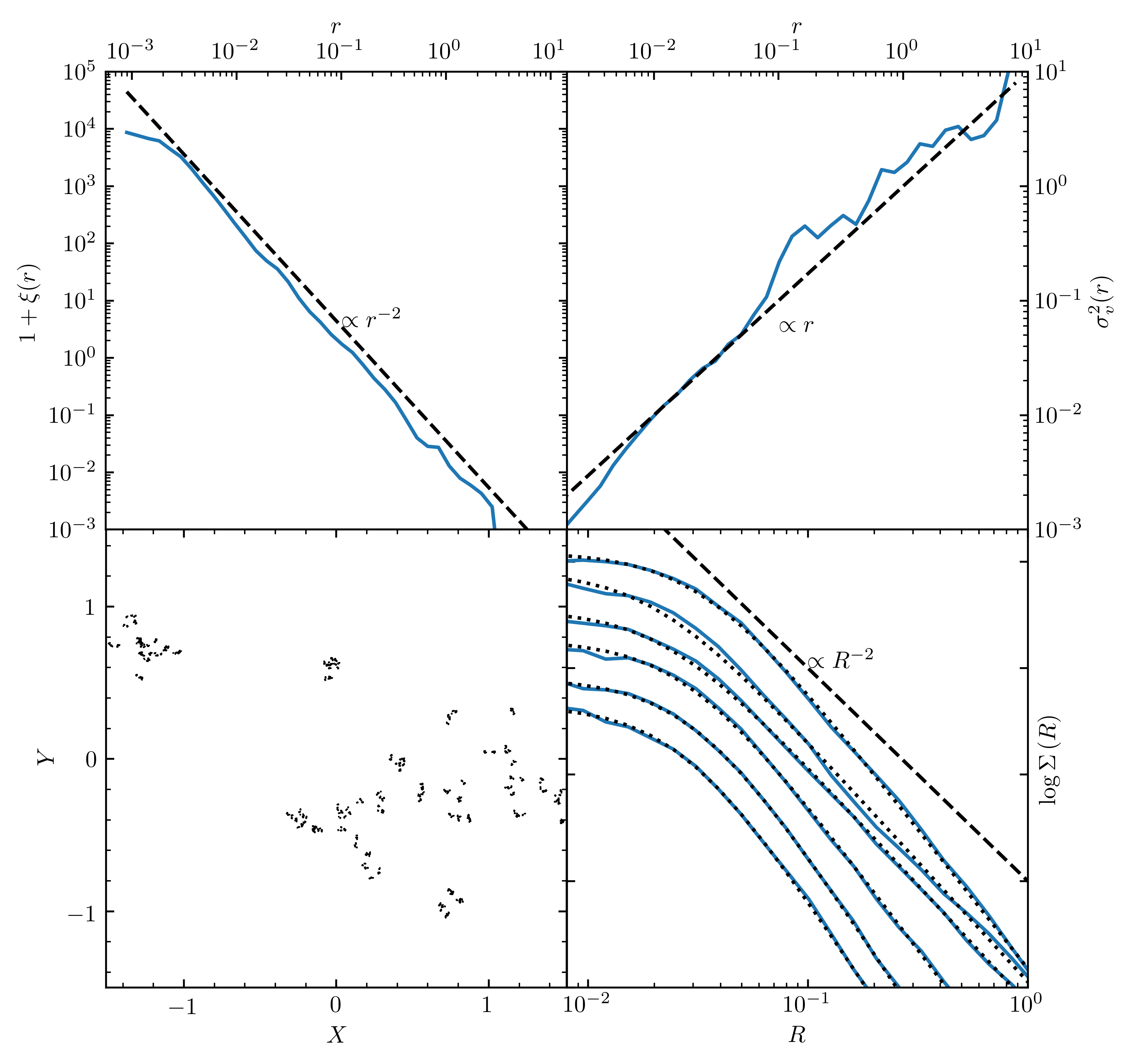}
\caption{Initial conditions and final results of a simulation of hierarchical cluster formation, as described in Section \ref{sec:Nbody:hierarchical}. {\it Top left:} Initial 3D correlation function of particle positions, which is $\propto r^{-2}$ above the resolution limit . {\it Top right:} Initial size-velocity dispersion relation. $\sigma_{v}^2\left(r\right)$ is the average velocity dispersion of particles within distance $r$ of any given point, and is constructed to be $\propto r$ to agree the observed relation of GMC kinematics \citep{bolatto:2008.gmc.properties}. {\it Lower left:} Initial hierarchically-clustered distribution of $64^3$ equal-mass particles, constructed by the stochastic fragmentation iteration described in Section \ref{sec:Nbody:hierarchical.ics}. {\it Lower right:} Surface density profiles of the best-resolved clusters formed by the end of the simulation. The profiles are offset from each other on the plot for visibility. They are well-described by the  \citetalias{Elson:1987.ymc.profile} model (Equation \ref{eq:EFF}).}
\label{fig:hierarchical.ics}
\end{figure*}

This procedure generates a clustered particle distribution with the desired spatial and velocity correlations, as shown in Figure \ref{fig:hierarchical.ics}. The gravitational binding energy $W$ for this distribution is computed with $G=1$ and the system is rescaled by a scale factor $\frac{1}{W}$ so that it has unit binding energy. The velocities are scaled to have a total kinetic energy of $0.5$, so that the system as a whole has a virial parameter $\alpha=\frac{T}{W}=0.5$.

\subsubsection{Results}
\begin{table}
\begin{tabular}{c|c|c|c|c|c}
\hline
\centering
Run & Mass & $R_{eff}$ & $\gamma$ & Ellipticity & $\hat{\chi}^2_{fit}$ \\ \hline
1 &   0.204 &   0.108 & $   2.62 \pm    0.02$ &  0.13 &    8.44 \\ 
1 &   0.202 &   0.166 & $   2.26 \pm    0.03$ &  0.17 &   39.87 \\ 
1 &   0.195 &   0.192 & $   2.23 \pm    0.02$ &  0.12 &   13.80 \\ 
1 &   0.115 &   0.074 & $   2.75 \pm    0.03$ &  0.15 &    4.46 \\ 
1 &   0.110 &   0.068 & $   3.16 \pm    0.02$ &  0.22 &    1.12 \\ 
1 &   0.054 &   0.052 & $   3.11 \pm    0.04$ &  0.16 &    1.71 \\ 
1 &   0.022 &   0.038 & $   3.15 \pm    0.05$ &  0.12 &    1.29 \\ 
1 &   0.019 &   0.035 & $   3.19 \pm    0.06$ &  0.15 &    1.31 \\ \hline
2 &   0.382 &   0.249 & $   2.28 \pm    0.03$ &  0.12 &  104.24 \\ 
2 &   0.364 &   0.171 & $   2.35 \pm    0.04$ &  0.17 &  105.77 \\ 
2 &   0.174 &   0.089 & $   2.89 \pm    0.03$ &  0.10 &    1.78 \\ \hline
3 &   0.147 &   0.099 & $   2.58 \pm    0.02$ &  0.13 &    4.36 \\ 
3 &   0.139 &   0.083 & $   2.75 \pm    0.03$ &  0.21 &    2.49 \\ 
3 &   0.114 &   0.078 & $   2.62 \pm    0.03$ &  0.20 &    9.19 \\ 
3 &   0.106 &   0.068 & $   2.78 \pm    0.03$ &  0.15 &    6.39 \\ 
3 &   0.092 &   0.067 & $   2.86 \pm    0.03$ &  0.26 &    2.96 \\ 
3 &   0.092 &   0.062 & $   3.22 \pm    0.07$ &  0.12 &    1.54 \\ 
3 &   0.053 &   0.050 & $   3.22 \pm    0.04$ &  0.33 &    1.33 \\ 
3 &   0.048 &   0.051 & $   3.17 \pm    0.07$ &  0.25 &    2.05 \\ 
3 &   0.045 &   0.046 & $   3.48 \pm    0.05$ &  0.20 &    1.28 \\ 
3 &   0.043 &   0.056 & $   2.76 \pm    0.03$ &  0.20 &    2.60 \\ 
3 &   0.031 &   0.040 & $   3.40 \pm    0.06$ &  0.13 &    1.63 \\ 
3 &   0.025 &   0.038 & $   3.32 \pm    0.05$ &  0.22 &    1.10 \\
\end{tabular}
\caption{Parameters of the clusters produced in the hierarchical relaxation simulations of Section \ref{sec:Nbody:hierarchical}: Masse, half-mass radius $R_{eff}$, fitted profile slope $\gamma$, ellipticity, and the reduced $\chi^2$ of the surface density fit to obtain $\gamma$. Uncertainties in $\gamma$ include the variation in the parameters from fitting the surface density profiles as projected in three different orthogonal planes.}
\label{table1}
\end{table}

We generate three different sets of initial conditions and evolve each system for 35 time units; the unit of time is on the order of the dynamical timescale of the system\footnote{A visualization of of Run 2 can be found at \url{http://www.tapir.caltech.edu/~mgrudich/hierarchical.mp4}}. Within the first few time units, sub-clusters undergo hierarchical assembly into a population of clusters that fly apart from each other and relax into a steady state. The rate-limiting step for the formation of a given cluster is merging timescale of its last two remaining sub-clusters, which is on the order of their mutual orbital period, at most on the order of several time units.

We identify bound clusters at the end of the simulation via the algorithm described in Appendix \ref{appendix:clusterfinding}. In general, roughly 80\% of particles are found to be gravitationally bound to a cluster, the rest having been dynamically ejected from their original hosts in the violent merging process. The surface density profiles of the clusters are generally well-fit by the \citetalias{Elson:1987.ymc.profile} model, and we present the fitted $\gamma$ values in Table \ref{table1}. The uncertainties quoted in Table \ref{table1} include the variation in the $\gamma$ obtained when projecting the surface density profile in three different orthogonal planes. This variation is generally small compared to the magnitude of $\gamma$, as the clusters are only weakly triaxial: their histories of statistically-isotropic mergers tend to average away preferred orientations. This is also reflected in the clusters' modest ellipticities, which we also tabulate in Table \ref{table1}. The ellipticities  lie in a similar range to those observed in the LMC cluster population \citep{1982MNRAS.199..565F,kontizas:1989.ymc.ellipticity}.

It is readily seen from Table \ref{table1} that the most massive clusters tend to have $\gamma$ closer to 2. The initial conditions were smoothed over an effective fixed mass scale $M_0$, so a hierarchically-assembled cluster of mass $M$ would have to have experienced an effective number of mergers $N=\log_2 \frac{M}{M_0}$, so in these simulations the more massive clusters have experienced more mergers, each of which creates a shallower profile. This anticorrelation between mass and $\gamma$ should not be interpreted as a prediction of the statistics of actual YMC populations, because observed YMCs are the product of many statistically-independent star formation events involving physics with only weak dependence on the mass scale (e.g. \citealt{fall:2010.sf.eff.vs.surfacedensity,guszejnov:2017.universal.scaling}). In contrast, we have simulated only three different events, all at a single mass scale.
 

In summary, these numerical experiments demonstrate that an EFF profile can emerge from the relaxation of a generic, hierarchically-clustered mass distribution with power-law spatial and kinematic scaling relations consistent with observations of GMCs and young star clusters.

\section{Discussion} \label{sec:discussion}


\subsection{Smooth vs. clumpy initial conditions for globular cluster formation}
\citet{goodwin:1998} concluded that the assembly of a YMC from an initially clumpy and asymmetric configuration was unlikely, for two main reasons. First, it was found that if the level of initial clumpiness is too great, some subclusters can survive for many orbits around the primary assembled cluster. However, \citet{goodwin:1998} simulated the evolution of a collection of clumps with comparable mass and uncorrelated initial positions, not accounting for correlations between subcluster positions imprinted by the structure formation process. This problem is averted by a hierarchical configuration, as neighbouring subclusters are all but guaranteed to merge. In the numerical experiments of Section \ref{sec:Nbody:hierarchical}, no persistent satellite clumps were found; the clusters that form tend to do so within a few dynamical times and disperse from each other, and within those clusters substructure is erased efficiently.

The other problem with clumpy initial conditions noted by \citet{goodwin:1998} was that the ellipticity of the final cluster is sensitive to the flattening of the initial conditions, and essentially any amount of initial flattening produced clusters with ellipticities much larger than have been observed,in the range $[0,0.28]$ \citep{kontizas:1989.ymc.ellipticity}. This problem is averted by the specific hierarchical picture we have considered in this work, wherein mergers at different levels in the hierarchy are uncorrelated in orientation due to an assumed statistical isotropy. From these experiments we find no cluster with ellipticity greatly exceeding the maximum observed. However, it should be noted that the assumption of statistical isotropy would not necessarily hold if, for example, the initial subclusters consisted of ``beads'' along a filament or a galactic spur. Indeed, it is quite possible that hierarchical star formation does impose large-scale statistical anisotropies. As such, an interesting direction for future work on this problem would be to investigate the effect of physically- or observationally-motivated anisotropy on hierarchical star cluster assembly. One avenue would be a straightforward modification to our fragmentation model (\S \ref{sec:Nbody:hierarchical.ics}) wherein the directions of the separations $\Delta x$ and relative velocities $\Delta v$ from one level to another are given a non-zero correlation.


Overall, we find the structure of YMCs to be largely compatible with the paradigm of hierarchical cluster formation that we have considered here. The constraints of \citet{goodwin:1998} upon clumpy initial substructure apply to the specific scenario that they simulated, with initial clumps of comparable masses and uncorrelated positions. The nature of the relaxation process appears to be qualitatively different when the initial stellar density and velocity field are initialized in a hierarchical fashion in the manner we have investigated, which takes into account the underlying spatial and kinematic correlation functions observed in star-forming regions.

\subsection{Applicability of the collisionless approximation}
Throughout this paper we have approximated the dynamics of the ensemble of stars by assuming that the evolution is collisionless and that stars of different masses are well-mixed. Working in this approximation, our $N$-body simulations represented the stellar distribution as an ensemble of equal-mass, gravitationally-softened particles. This picture is clearly not entirely realistic for star clusters, which are generally are dense enough for stellar close encounters to be common enough to affect their long-term dynamical evolution. \citet{bonnell:2003.hierarchical} found that an order-unity fraction of stars have close encounters during hierarchical star cluster formation, so the the granularity of stellar mass should clearly have some effect. We expect the collisionless approximation to break down for clusters in which the the 2-body relaxation time is less than the orbital time, which Equation \ref{eq:relaxation} predicts is the case for clusters less massive than $\sim 250\msun$. Therefore, we expect the physics considered in this work to be most applicable to the regime of massive star clusters that assembled from sub-clusters more massive than this.

The success of the collisionless approximation in producing star clusters with realistic coarse-grained structure in both multi-physics star cluster formation simulations \citepalias{grudic:2016.sfe} and the numerical experiments of this paper suggests that it is be sufficient for these purposes. The orbital evolution in the hierarchical merging scenario is dominated by rapid changes in the gravitational potential driving violent relaxation, which affects stellar trajectories independently of their mass \citep{lynden-bell67}. 

\subsection{Star cluster initial conditions}
It has become possible in recent years to simulate the direct $N$-body evolution, and other processes governing the post-formation dynamical evolution, of a globular cluster consisting of as many as $\sim 10^6$ stars \citep{wang:2016.dragon}. Such simulations are important for understanding the rich variety of physical mechanisms that caused young star clusters to evolve into present-day mature globulars, but they must assume some initial cluster properties ad-hoc. Typically, either the \citet{plummer} or \citet{king:profile} model is used as the initial model \citep{portegies-zwart:2010.starcluster.review}.

However, since YMCs are well-described by the  \citetalias{Elson:1987.ymc.profile} model, and we have given this observation further physical motivation in this paper, we propose that a shallower \citetalias{Elson:1987.ymc.profile} model is a more realistic initial condition for globular cluster simulations, rather than something that resembles a mature globular cluster. According to the distribution of profile slopes (Figure \ref{fig:gamma_cdf}), a typical model would have $\gamma\sim2.5$. Compared to a Plummer model of equal mass and half-mass radius, the central density of a $\gamma=2.5$ profile is more than ten times greater, so collisional effects such as mass segregation and core collapse would likely have much earlier onset \footnote{Although they would still take longer than the initial formation of the cluster.}. This could easily mark the difference between runaway core collapse happening before or after the mass loss and death of massive stars $\sim\unit[3]{Myr}$ after star formation. This is a critical factor determining whether it is possible for runaway stellar mergers to form a very massive star or an IMBH in the centre of the cluster \citep{portegies-zwart:2002,gurkan2004,freitag:2006.runaway.collisions}. It should also influence the pairing and hardening of massive stellar binaries centre of dense clusters, which would alter the rate of massive (e.g. $ \sim \unit[60]{\msun}$) binary black hole mergers like GW150914 \citep{GW150914, rodriguez:2015.bbh.globulars, rodriguez:2016.bbh.globulars}. Clearly the detailed early dynamical evolution of realistic YMC models warrants further study with more realistic initial conditions.

\subsection{The outer NFW profile}
We have established that the phase-space dilution caused by violent relaxation and phase mixing in the hierarchical merging of star clusters generally drives clusters toward shallower mass profiles approaching $\rho \propto r^{-3}$. Cold dark matter halos also merge hierarchically, and are generally well-described by the \citet{nfw:profile} (NFW) profile in cosmological simulations, which also has an $r^{-3}$ dependence. Indeed, it has long been established that such a profile has some relationship with hierarchical merging \citep{white:1979.mergers,villumsen:1982,duncan:1983,mcglynn:1984,pearce:1993}. To explain this, we cannot invoke exactly the same argument as the one we have made for star clusters in \S\ref{sec:analytic}, because the NFW model has no maximum phase-space density to conserve. Nevertheless, the \citet{dehnen:2005.mixing} mixing theorem still implies that the hierarchical merging of dark matter halos cannot create steeper density profiles. Furthermore, the outer density profile should behave in a manner that is insensitive to the details of whether the inner profile is a core or a cusp, so shallower density profiles should generally result in mergers. We therefore argue that the $\propto r^{-3}$ outer NFW profile can be understood as the endpoint of the same process of phase-space dilution that we have argued drives star clusters to shallow density profiles.  

\section{Conclusions}
\label{sec:conclusion}
We arrive at the following conclusions about the formation of young massive clusters: 
\begin{itemize}
\item We compile observational data of young massive cluster populations \citep{ryon:2015.m83.clusters,ryon:2017.ymc.profiles, mackey:2003.ymc.profiles.lmc,mackey:2003.ymc.profiles.smc} and find that the distribution of surface brightness profile slopes (Figure \ref{fig:gamma_cdf}) is similar between different cluster populations, suggesting that it is universal due to common star formation physics.
\item MHD star cluster formation simulations with resolved cooling, fragmentation, and stellar feedback \citep{grudic:2016.sfe} have produced a population of star clusters with profile slopes that agree with observations (Figure \ref{fig:gamma_cdf}), despite the fact that the simulations do not resolve the formation of individual stars. To capture the essential physics that determine the shapes of nascent massive star clusters, it suffices to resolve some fraction of the dynamic range of fragmentation. 
\item Stellar feedback clearly has an important role in shaping star clusters, as simulations without feedback are different from observed YMCs in many ways. The role of stellar feedback in setting star cluster structure should be elucidated in detailed cluster formation simulations.
\item Based on the the observational and simulation data mentioned above, evidence is strong that a YMC's profile slope is established when it is dynamically young, so must be established in the cluster formation process.
\item We develop an analytic model for the evolution of a cluster's profile slope $\gamma$ in a sequence of collisionless pairwise mergers between star clusters modelled by the \citetalias{Elson:1987.ymc.profile} model. Phase-space mixing requires that the final slope is no shallower than that of either progenitor. Furthermore, assuming conservation of mass, energy, and maximum phase-space density, we find that mergers must always shallow the slope toward 2 by some amount. Thus a sufficiently large number of hierarchical mergers will result in $\gamma \sim 2$, as argued in \cite{guszejnov:2017.universal.scaling} from more general considerations.
\item We perform collisionless $N$-body simulations of three iterated star cluster mergers, starting with a pair of identical \citet{plummer} models and then merging the result with a copy of itself twice. The results of these simulations are in good agreement with our analytic model: at most $\sim10\%$ of mass and energy are ejected in each merger, the maximum phase-space density is conserved, and the mergers drive $\gamma$ from 4 initially to a value close to 2 (Table 1). The collisionless merger of two \citetalias{Elson:1987.ymc.profile} clusters produces another cluster whose surface density profile is also well-described by the \citetalias{Elson:1987.ymc.profile} model, however deviations from the model are more apparent in the phase-space structure (Figure \ref{fig:pairwise.merger}).
\item We have performed $N$-body experiments following the collisionless relaxation of a hierarchically-clustered mass distribution with spatial and kinematic scaling relations corresponding to those observed in GMCs and young star clusters. We find that sub-clusters rapidly merge hierarchically into steady-state star clusters with \citetalias{Elson:1987.ymc.profile}-like surface density profiles, despite no initial surface density model being assumed. Thus the \citetalias{Elson:1987.ymc.profile} model is physically motivated within the paradigm of hierarchical star cluster formation, and indeed \citetalias{Elson:1987.ymc.profile}'s explanation in terms of dissipationless relaxation following rapid star formation is venerated. 
\item Because clusters resembling YMCs emerge so readily from plausible star formation physics, a shallow EFF profile is a more plausible model of a nascent star cluster than the commonly-simulated \citet{plummer} or \citet{king:profile} models. This may have interesting implications for the detailed dynamical evolution of dense star clusters.
\end{itemize}
\section*{Acknowledgements}
We thank Michael S. Fall, Bruce Elmegreen, Scott Tremaine, and the anonymous referee for helpful feedback. Support for MG and TAPIR co-authors was provided by an Alfred P. Sloan Research Fellowship, NASA ATP Grant NNX14AH35G, and NSF Collaborative Research Grant \#1411920 and CAREER grant \#1455342. MB-K acknowledges support from NSF grant AST-1517226 and from NASA grants NNX17AG29G and HST-AR-13888, HST-AR-13896, and HST-AR-14282 from the Space Telescope Science Institute, which is operated by AURA, Inc., under NASA contract NAS5-26555. Numerical calculations were run on the Caltech compute clusters `Zwicky' (NSF MRI award $\#$ PHY-0960291) and `Wheeler' and allocation TG-AST130039 granted by the Extreme Science and Engineering Discovery Environment (XSEDE) supported by the NSF.



\bibliographystyle{mnras}
\bibliography{master} 



\appendix
\section{Cluster finding algorithm}
\label{appendix:clusterfinding}
To identify bound star clusters from the star particle mass, velocity and position data of the \citetalias{grudic:2016.sfe} simulations, we use an algorithm based on identifying potential wells. This is generally more robust than methods based on identifying density maxima because the gravitational potential contains all necessary information for cluster finding, while being inherently smoother and hence less susceptible to noise. The algorithm is as follows:
\begin{enumerate}
\item Determine some fixed number $N_{ngb}$ of each star particle's nearest neighbors in position space.
\item From each particle, move to the neighbour particle with the lowest gravitational potential. Repeat until a local minimum in the potential is found. This is the bottom of the potential well, to which the initial particle is now ``associated''.
\item Compute the gravitational potential as sourced only by the particles associated with this potential well in isolation.
\item Associated particles that are bound to the potential well are considered bound members of the cluster.
\end{enumerate}

In practice, we take $N_{ngb}=32$, which is the number of neighbour elements used for constructing the hydrodynamic mesh and force softening in the simulations, so it is on the order the size of the least massive self-gravitating structure that can exist in the simulation. A larger value could potentially lump together distinct bound star clusters, while smaller values generally increase the population of spurious clusters. We find this algorithm to have satisfactory accuracy for this problem; it has been tested on control datasets for which the cluster associations are known a priori, and stably identifies the same cluster between different simulation snapshots.

\section{(Semi-) Analytic Properties of the EFF model} \label{appendix:EFF}
Here we derive useful quantities for calculations involving star clusters modeled by the  \citetalias{Elson:1987.ymc.profile} density profile (Equation \ref{eq:EFF3D}) with arbitrary profile slope $\gamma$:
\begin{equation}
\rho\left(r\right) = \rho_0 \left(1 + \frac{r^2}{a^2}\right)^{-\frac{\gamma+1}{2}}.
\end{equation}
The quantities needed to construct a dynamical model with this density profile are only generally expressible in closed form in the special case $\gamma=4$, which is the \citet{plummer} model. This has ensured its popularity as an initial condition for $N$-body simulations that is easy to construct. However, as discussed in Section \ref{sec:gammadist}, a much more typical initial condition for a star cluster would be $\gamma\sim 2-3$. For those quantities that lack closed-form expressions, we provide approximate expressions or upper and lower bounds for use with numerical root solvers. The reader is also directed to \citep{1989ApJ...347..201L} for the derivation of the collisionless Jeans model.

\subsection{Cumulative mass distribution}
The cumulative mass distribution for arbitrary $\gamma$ is:
\begin{align}
M\left(<r \right) &= \int_0^r 4 \pi r'^2 \rho\left(r'\right) \mathrm{d} r' \nonumber \\
&=\frac{4 \pi \rho_0}{3} r^3 \, _2F_1\left(\frac{3}{2},\frac{\gamma+1}{2};\frac{5}{2};-\frac{r^2}{a^2}\right), 
\end{align}
where $_2F_1\left(a,b;c;z\right)$ is the Gauss hypergeometric function \citep[chap. 15]{abramowitz.stegun}.
\subsection{Half-mass radius}
\begin{figure}
\includegraphics[width=\columnwidth]{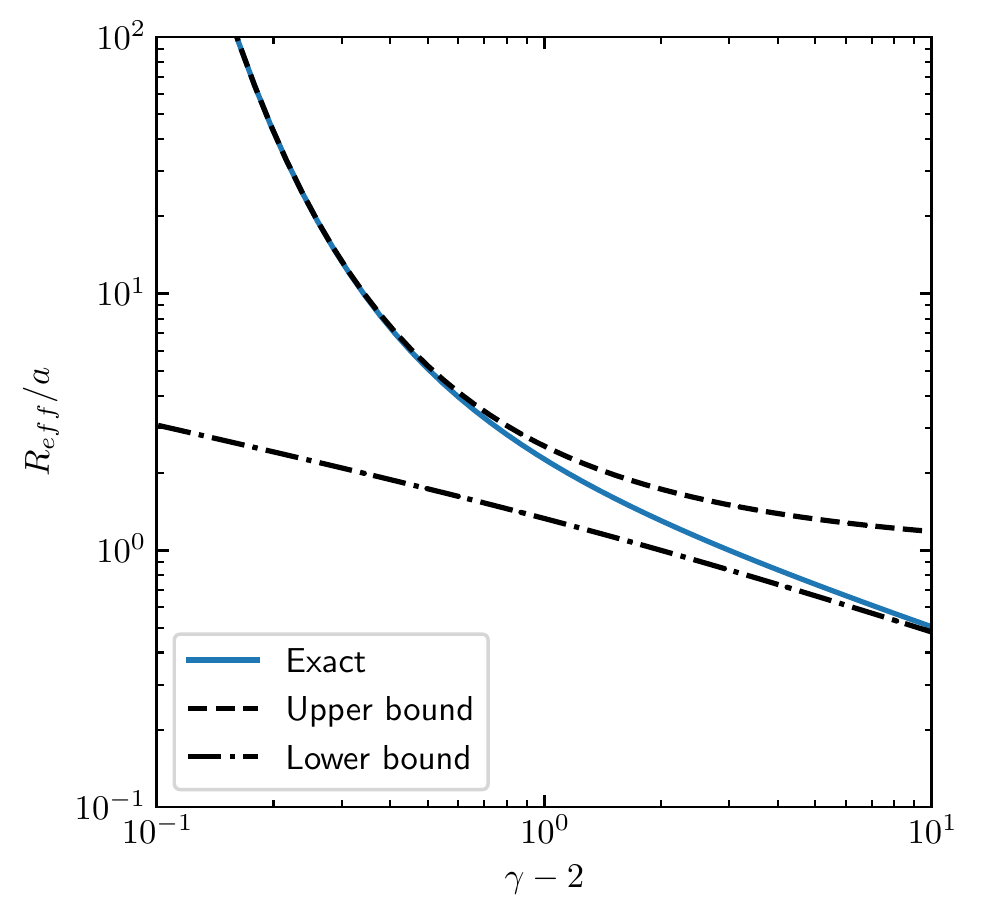}
\caption{Three-dimensional half-mass radius $R_{eff}$ as a function of $\gamma$ in units of the scale radius $a$. The numerical solution is shown in blue, between the bounds given in Equation \ref{reff}.}
\end{figure}
The three-dimensional half-mass radius $R_{eff}$ may be obtained by solving $M\left(<r\right)/M=\frac{1}{2}$. For the Plummer model ($\gamma=4$), the solution is $\frac{1+\sqrt[3]{2}}{\sqrt{3}}a \approx 1.3$. For general $\gamma$, there is no closed form solution. We may derive upper and lower bounds from the constant and power-law parts of the density profile respectively from the expansions of $M\left(r\right)$ about 0 and $\infty$:

\begin{equation}
\left(\frac{3 M }{4 \pi \rho_0}\right)^\frac{1}{3} \leq R_{eff}\leq \left(\frac{4\Gamma \left(\frac{\gamma +1}{2}\right)}{\sqrt{\pi} \Gamma \left(\frac{\gamma
   }{2}\right)}\right)^{\frac{1}{\gamma -2}} a.
   \label{reff}
\end{equation}

Equipped with these bounds, $R_{eff}$ can be computed efficiently with a bounded root-finding algorithm such as Brent's method. In the limit $\gamma \rightarrow 2$, the solution will approach the upper bound, as most of the mass will be in the power-law portion. Similarly $R_{eff} \rightarrow \left(\frac{3 M }{4 \pi \rho_0}\right)^\frac{1}{3}$ as $\gamma \rightarrow \infty$ because most of the mass will be in the core.

\subsection{Potential}
The gravitational potential is given by the integral
\begin{align}
\Phi \left(r\right) &= \int_\infty ^r \frac{G M\left(r'\right)}{r'^2} \mathrm{d}r' \nonumber \\
&= -\frac{4 \pi G a^2 \rho _0 \, _2F_1\left(\frac{1}{2},\frac{\gamma
   -1}{2};\frac{3}{2};-\frac{r^2}{a^2}\right)}{\gamma -1}. 
\label{potential}
\end{align}
The expansion of $\Phi\left(r\right)$ about the center is:
\begin{equation}
\Phi\left(r\right) = 4 \pi G \rho_0\left(\frac{r^2}{6} - \frac{a^2}{\gamma-1}\right) + \mathcal{O}\left(r^4\right).
\end{equation}
The shortest possible orbital frequency in the cluster is that associated with simple harmonic motion in the central potential well, which depends only on the central density:
\begin{equation}
\Omega_{max} = \sqrt{\frac{4\pi G \rho_0}{3}}.
\end{equation}

Expanding about $r=\infty$, we see that the leading order correction to the monopole term $-\frac{GM}{r}$ is:
\begin{equation}
\Phi\left(r\right) + \frac{GM}{r} \approx \frac{GM}{a} \frac{\Gamma \left(\frac{\gamma -1}{2}\right)}{\sqrt{\pi } \Gamma \left(\frac{\gamma }{2}\right)} \left(\frac{r}{a}\right)^{1-\gamma }.
\end{equation}

Thus, for larger values of $\gamma$, the leading correction to the point mass potential is $\propto r^{1-\gamma}$, which will be very small, so the potential is well-approximated by a Keplerian potential. This approximation will be less valid for $\gamma\rightarrow 2$, as most of the mass will be in the power law portion of the profile.

\subsection{Energy}
\begin{figure}
\includegraphics[width=\columnwidth]{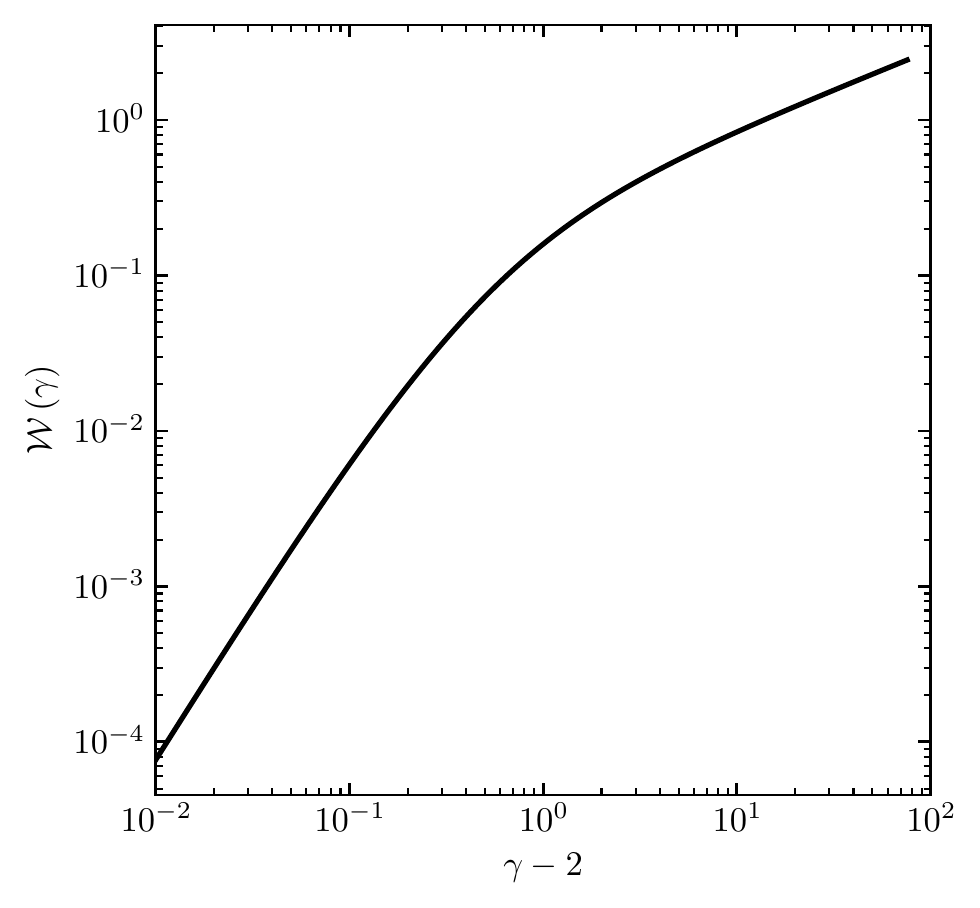}
\caption{$\mathcal{W}\left(\gamma\right)$ as a function of $\gamma$, where the gravitational binding energy is given by $W=\mathcal{W}\left(\gamma\right) \frac{G M^2}{a}$. The function is very well approximated by Equation \ref{Wapprox}. It is $\propto \left(\gamma-2\right)^2$ in the limit $\gamma \rightarrow 2$ and $\propto \left(\gamma-2\right)^\frac{1}{2}$ in the limit $\gamma \rightarrow \infty$.}
\label{fig:energy}
\end{figure}

A star cluster in dynamical equilibrium will satisfy the virial theorem: $E=-W/2$, where $W$ is the magnitude of the gravitational potential energy. The potential energy associated with the mass distribution may be computed as the integral:
\begin{equation}
W = \int_0^\infty \frac{G M\left(r\right)}{r} 4 \pi r^2 \rho\left(r\right) \mathrm{d}r = \mathcal{W}\left(\gamma\right) \frac{G M^2}{a},
\end{equation} 
where $\mathcal{W}\left(\gamma\right)$ is a dimensionless function of $\gamma$, plotted in Figure \ref{fig:energy}. For the Plummer model, $\mathcal{W}\left(\gamma\right)=\frac{3\pi}{32}$. The expression in terms of the hypergeometric function is cumbersome, however it is asymptotically $\propto \left(\gamma-2\right)^2$ as $\gamma\rightarrow2$ and $\propto \left(\gamma-2\right)^\frac{1}{2}$ as $\gamma\rightarrow \infty$. It can be very well approximated by the following expression:
\begin{equation}
\mathcal{W}\left(\gamma\right) = \left(\left(c_1 (\gamma -2)^2\right)^\alpha + \left(c_2 \left(\gamma-2\right)^\frac{1}{2}\right)^\alpha \right)^\frac{1}{\alpha},
\label{Wapprox}
\end{equation}
with $c_1=0.780$, $c_2=0.284$, and $\alpha = -0.692$. This expression interpolates between the two asymptotic behaviours, and is indistinguishable from $\mathcal{W}\left(\gamma\right)$ as plotted in Figure \ref{fig:energy}.

\subsection{Phase-Space Distribution Function}
\begin{figure}
\includegraphics[width=\columnwidth]{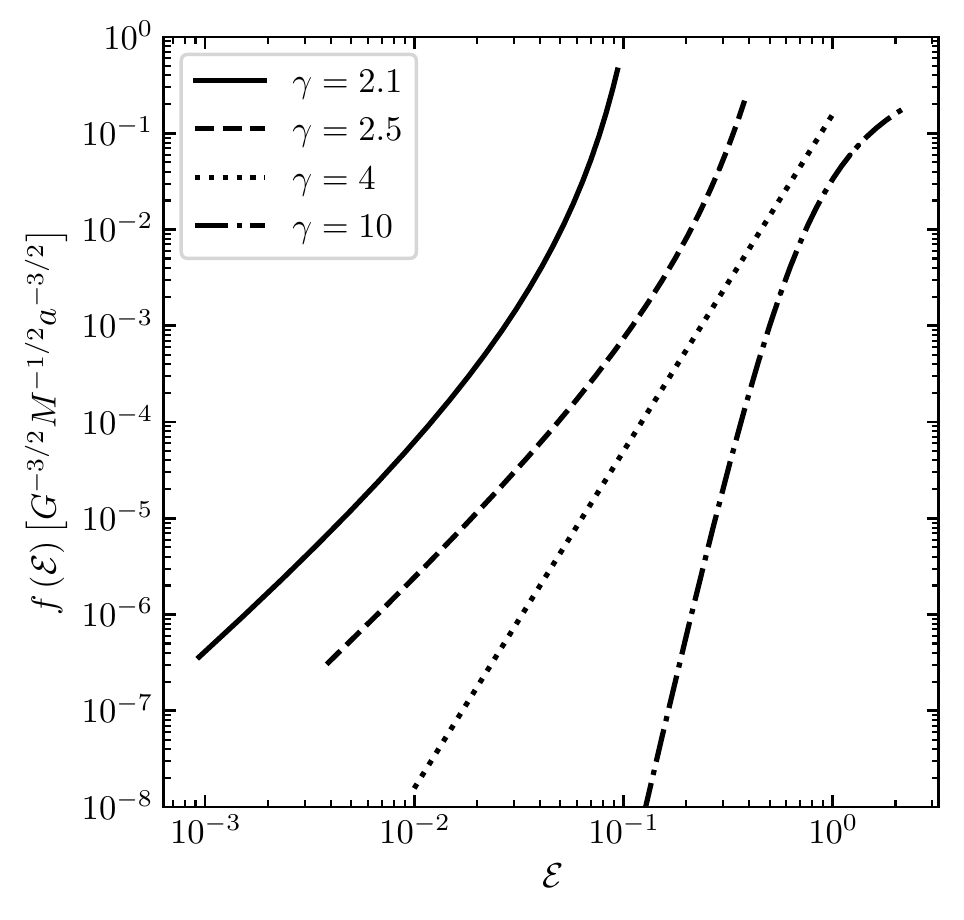}
\caption{Phase-space density $f\left(\mathcal{E}\right)$ in units of $G^{-3/2} M^{-1/2} a^{-3/2}$ for isotropic cluster models with different $\gamma$. The Plummer model ($\gamma=4$) is the only one that is a true power law $\propto \mathcal{E}^{7/2}$, hence its popularity as an analytic model for $N$-body initial conditions.}
\label{fe}
\end{figure}

\begin{figure}
\includegraphics[width=\columnwidth]{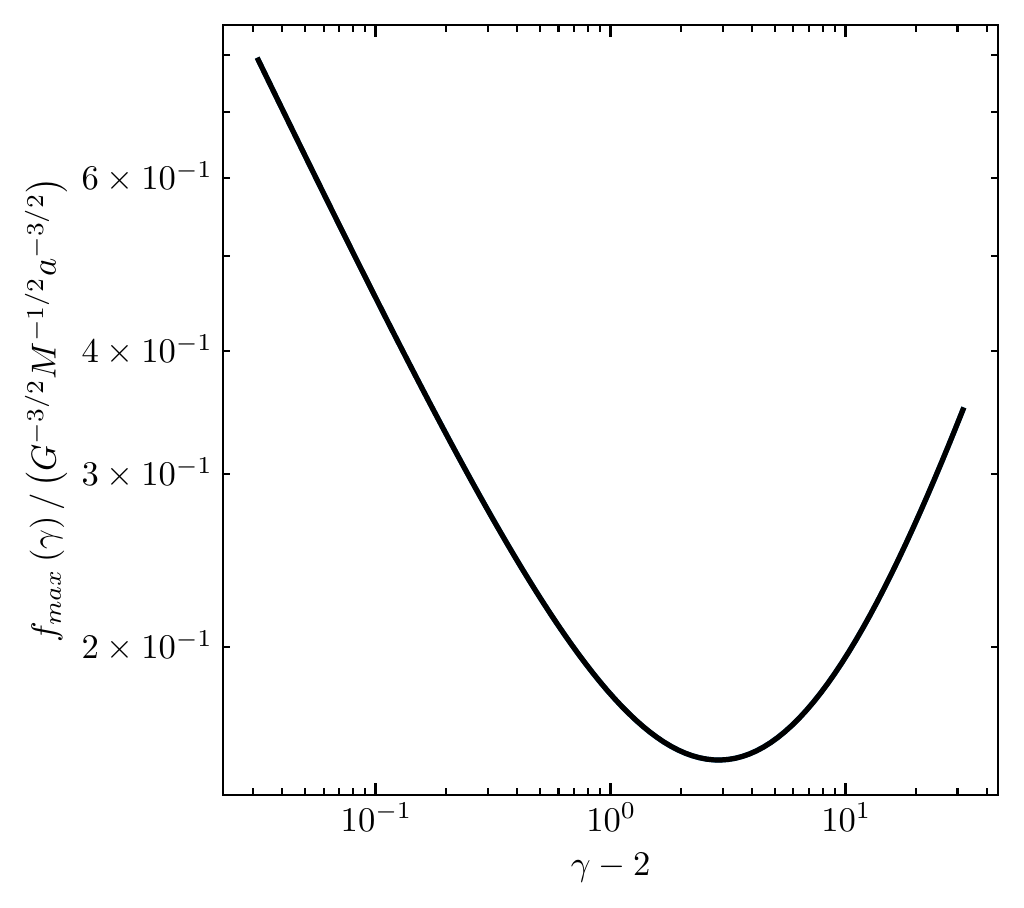}
\caption{Maximum phase-space density $f_{max}$ as a function of $\gamma$, in units of $G^{-3/2} M^{-1/2} a^{-3/2}$. The function is $\propto \left(\gamma-2\right)^{-1/2}$ in the limit $\gamma\rightarrow 2$, $\propto \left(\gamma-2\right)^{3/4}$ in the limit $\gamma\rightarrow \infty$, and minimized for the Plummer model ($\gamma=4$). It is well approximated by Equation \ref{Fapprox}.}
\label{fig:fmax}
\end{figure}
With the potential given by Equation \ref{potential}, and assuming an isotropic velocity distribution, the phase-space density $f\left(\textit{\textbf{x}},\textit{\textbf{v}}\right)$ is a function of specific orbital energy alone. We may determine the phase-space density $f\left(\mathcal{E}\right)$ with the usual integral formula \citep{binneytremaine}:
\begin{equation}
f\left(\mathcal{E}\right) = \frac{1}{\sqrt{8}\pi^2} \frac{\mathrm{d}}{\mathrm{d}\mathcal{E}} \int_{\psi=0}^{\psi=\mathcal{E}} \frac{\mathrm{d}\rho}{\sqrt{\mathcal{E}-\psi}},
\label{phasespaceint}
\end{equation}
where $\psi=-\Phi$ and $\mathcal{E}=\left(-\Phi - \frac{1}{2}v^2\right)$. In the limit $r >> R_{eff}$, we may approximate $f\left(\mathcal{E}\right)$ by substituting the Keplerian potential and the approximation $\rho \sim \rho_0 r^{-\gamma-1}$. In this limit:

\begin{equation}
f\left(\mathcal{E}\right) \approx \frac{ \Gamma (\gamma +1) \Gamma \left(\frac{\gamma +3}{2}\right)}{\sqrt{2} \pi ^3 \Gamma \left(\frac{\gamma -2}{2}\right) \Gamma \left(\gamma +\frac{1}{2}\right)} \mathcal{E}^{\gamma -\frac{1}{2}}
\label{eq:feapprox}
\end{equation}

Remarkably, for the Plummer model ($\gamma = 4$), this power law approximation holds exactly. For all other values this is not so, and the integral and derivative in Equation \ref{phasespaceint} must be taken numerically. The derivative in Equation \ref{phasespaceint} may be taken with a high-order finite difference stencil, as the integral is smooth everywhere except at $\mathcal{E}=\psi\left(0\right)$. Figure \ref{fe} plots the numerically-computed $f\left(\mathcal{E}\right)$ for various values of $\gamma$. It is clear that for the non-Plummer models, the phase-space distribution for the lowest energy (largest $\mathcal{E}$) orbits deviates significantly from a power law. Figure \ref{fig:fmax} shows the dependence of the maximum phase-space density upon $\gamma$. In the usual units in terms of $G$, $M$ and $a$, the Plummer model has the lowest maximum phase-space density, and with $M$ and $a$ held constant $f_{max}$ increases without bound as $\gamma\rightarrow2$ and $\gamma \rightarrow \infty$. We may roll the $\gamma$ dependence into a dimensionless function $\mathcal{F}\left(\gamma\right)$, such that $f_{max} = \mathcal{F}\left(\gamma\right) G^{-3/2}M^{-1/2} a^{-3/2}$. An approximation of $\mathcal{F}$ with maximum error $\sim 10^{-4}$ over $\gamma \in \left[2.01,10\right]$ is:
\begin{equation}
\mathcal{F}\left(\gamma\right) \approx \left(\left(c_1 (\gamma -2)^\frac{3}{4}\right)^\alpha + \left(c_2 \left(\gamma-2\right)^{-\frac{1}{2}}\right)^\alpha \right)^\frac{1}{\alpha},
\label{Fapprox}
\end{equation}
where $c_1 = 0.0228$, $c_2=0.139$, and $\alpha=0.816$.

\subsection{Cumulative Phase-Space-Density Distribution $M\left(<f\right)$}\label{appendix:EFF:mf}
\begin{figure}
\includegraphics[width=\columnwidth]{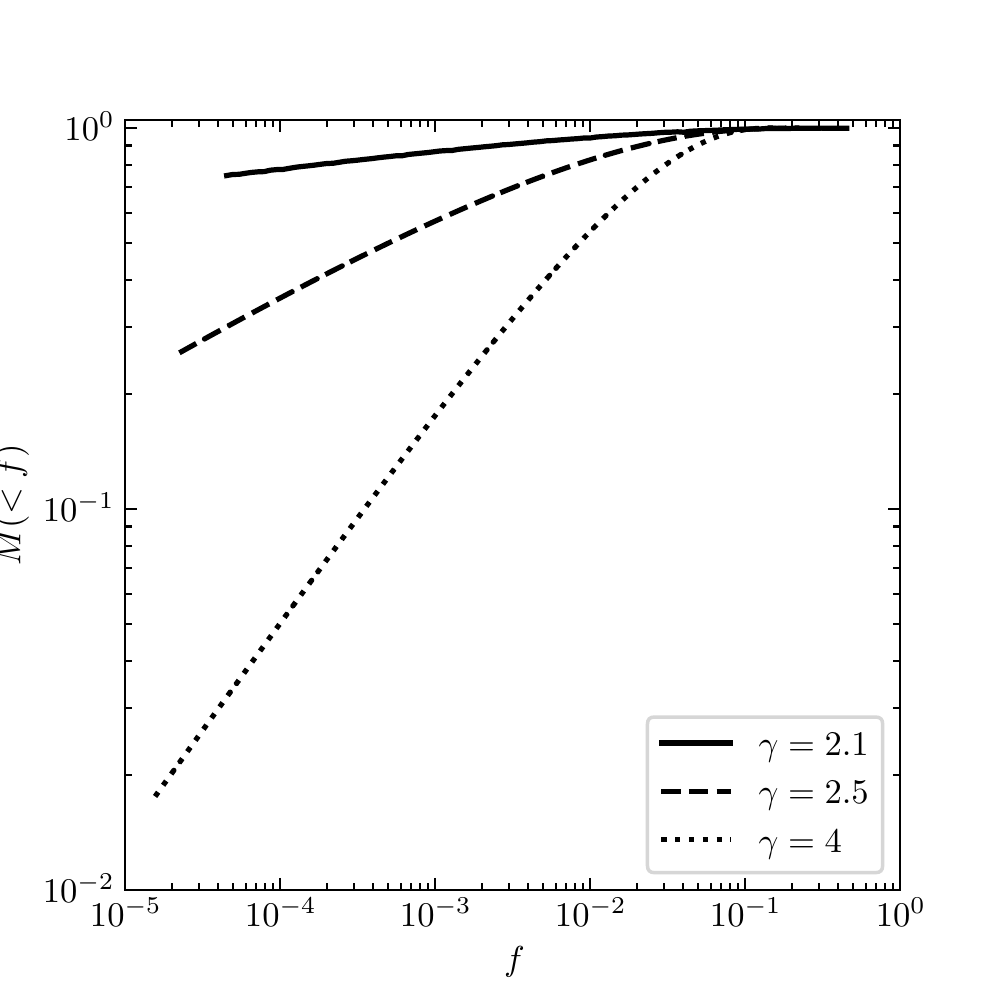}
\caption{Cumulative phase-space density distribution $M\left(<f\right)$ for a series of clusters varying $\gamma$ while keeping mass and energy fixed. At equal mass and energy, the distribution is more spread-out for $\gamma$ values closer to 2, and is asymptotically $\propto f^\frac{2\gamma-4}{2\gamma-1}$.}
\label{fig:mf}
\end{figure}


$M\left(<f\right)$, the amount of mass at phase-space density less than $f$, is a useful diagnostic quantity in $N$-body simulations because it is robust to noisy estimates of $f$ from Monte Carlo particle data. It is also useful for placing analytic constraints on merger products because it strictly increases in collisionless evolution as phase mixing occurs.

For a spherically-symmetric, isotropic cluster model, $f$ is a monotonic function of $\mathcal{E}$, so it is convenient to compute $M\left(<f\right)$ as the integral
\begin{equation}
M\left(<f\right) = \int_0^{\mathcal{E}(f)}f\left(\mathcal{E}\right) g\left(\mathcal{E}\right)\mathrm{d} \mathcal{E},
\label{eq:Mf}
\end{equation}
where $\mathcal{E}(f)$ is the inverse function of $f(\mathcal{E})$ and $g\left(\mathcal{E}\right)\mathrm{d}\mathcal{E}$ is the phase-space volume within the interval $\left[\mathcal{E},\mathcal{E}+\mathrm{d}\mathcal{E}\right]$, computable as
\begin{equation}
g\left(\mathcal{E}\right) = \sqrt{2}\left(4\pi\right)^2 \int_0 ^{r\left(\mathcal{E}\right)} r^2 \sqrt{\psi\left(r\right) - \mathcal{E}}\,\mathrm{d}r,
\end{equation}
where again $r\left(\mathcal{E}\right)$ is the radius at which $\psi\left(r\right)=\mathcal{E}$. In the Keplerian approximation, this gives
\begin{equation}
g\left(\mathcal{E}\right) \approx \sqrt{2 M a^5} \pi^3 \mathcal{E}^{-5/2}.
\label{geapprox}
\end{equation}
Combining this with \ref{eq:feapprox}, the asymptotic form of $M\left(<f\right)$ is
\begin{equation}
M\left(<f\right) \approx \frac{2^{\frac{\gamma -2}{2 \gamma -1}} \pi ^{\frac{9}{1-2 \gamma }+3} \hat{f} \left(\frac{\hat{f} \Gamma \left(\frac{\gamma }{2}-1\right) \Gamma \left(\gamma +\frac{1}{2}\right)}{\Gamma (\gamma +1) \Gamma \left(\frac{\gamma+3}{2}\right)}\right)^{\frac{3}{1-2 \gamma }}}{\gamma -2} \propto f ^ {\frac{2\gamma-4}{2\gamma-1}},
\end{equation}

where $\hat{f} = f / \left(G^{-3/2} M^{-1/2} a^{-3/2}\right)$. In general, the integral \ref{eq:Mf} must be performed numerically. In Figure \ref{fig:mf}, we plot $M\left(<f\right)$ for a sequence of  \citetalias{Elson:1987.ymc.profile} clusters with varying $\gamma$ but equal mass and energy. Note how smaller values of $\gamma$ have a flatter distribution, so their mass is effectively spread over more orders of magnitude in $f$.


\bsp	
\label{lastpage}
\end{document}